\documentclass[aps,prb,amsmath,amssymb,floatfix,twocolumn]{revtex4-1}
\pdfoutput=1
\usepackage{times}
\usepackage{bbold}
\usepackage{graphicx}
\usepackage{color}
\usepackage{hyperref}
\usepackage{wasysym}

\begin{document}
 
\title{Superconductivity from fractionalized excitations in  an enigmatic heavy fermion material}
\author{Chen-Hsuan Hsu}
\affiliation{Department of Physics and Astronomy, University of
California Los Angeles, Los Angeles, California 90095-1547}
\author{Sudip Chakravarty}
\affiliation{Department of Physics and Astronomy, University of
California Los Angeles, Los Angeles, California 90095-1547}

\date{\today}
 
\begin{abstract}
 
An unconventional pairing mechanism in the heavy-fermion material $\mathrm{URu_{2}Si_{2}}$ is studied. We propose a mixed singlet-triplet $d$-density wave to be the hidden-order state in $\mathrm{URu_{2}Si_{2}}$. The exotic order is topologically nontrivial and supports a charge $2e$ skyrmionic spin texture, which is assumed to fractionalize into merons and antimerons at the deconfined quantum critical point. The interaction between these fractional particles results in a (pseudo)spin-singlet chiral $d$-wave superconducting state, which breaks time reversal symmetry. Therefore, it is highly likely to produce a nonzero signal of the polar Kerr effect at the onset of the superconductivity, consistent with recent experiments. In addition, the nodal structures of the possible pairing functions in our model are consistent with the thermodynamic experiments in $\mathrm{URu_{2}Si_{2}}$.
 
\end{abstract}
 
 
\maketitle
 
\section{Introduction}
The identity of the order parameter in $\mathrm{URu_{2}Si_{2}}$ (URS),~\cite{Mydosh:2011} a heavy-fermion material, below the so called hidden-order (HO) transition at $T_{\textrm{HO}}=17.5$~K is unknown despite its discovery a quarter century ago.~\cite{Santini:1994,Kiss:2005,Haule:2009,Kusunose:2011,Cricchio:2009,Harima:2010,Ikeda:1998,Mineev:2005,Fujimoto:2011,Das:2012,Elgazzar:2009,Chandra:2002,Varma:2006,Chandra:2013,Rau:2012,Dubi:2011,Ikeda:2012,Pepin:2011,Riseborough:2012,Yuan:2012}
Buried deep inside this phase lies a much less explored unconventional  superconducting state at a temperature $T_{c}\sim 1.5$~K.~\cite{Palstra:1985,Maple:1986,Schlabitz:1986,deVisser:1986,Mydosh:2011}  It is natural that there must be an intimate relation between the two. While numerous theoretical models have been proposed to explain the HO phase, very few of them attempted to explain the origin of the unconventional superconductivity. Thus, it is our central interest to explore the connection between the two states to provide a mechanism for the unconventional superconducting state.

In this paper we posit that an intriguing density wave state, termed mixed singlet-triplet $d$-density wave (st-DDW),~\cite{Hsu:2011,Hsu:2013} is responsible for the HO state. This state has no net charge or spin modulations and does not break time reversal symmetry (TRS). It does have topological order with a quantized spin Hall effect.~\cite{Hsu:2011} Thus, it is naturally impervious to common experimental probes.~\cite{Hsu:2013} We then construct a global phase diagram in which there is a deconfined quantum critical point (QCP),~\cite{Senthil:2004} which is ultimately responsible for the basic mechanism of superconductivity. The skyrmionic spin texture in the st-DDW state fractionalizes into fermonic merons and antimerons,~\cite{Senthil:2004} which results in unconventional chiral $d$-wave superconductivity.~\cite{Kasahara:2007,Yano:2008} The deconfinement takes place only at the QCP. On one side of it merons and antimerons are paired to form skyrmions, but on the other side merons pairs with merons, and similarly for antimerons. The resulting superconducting state breaks TRS, which can be directly detected by polar Kerr effect (PKE) measurements.~\cite{Kapitulnik:2009} 
Determination of the density wave state posited here is also possible through two-magnon Raman scattering, nuclear quadrupolar resonance, or the skyrmions themselves. In a more general context, our work reflects the rich possibilities of emergent behavior in condensed matter systems. 

Density wave states of higher angular momenta are intriguing objects.~\cite{Nayak:2000,*Nersesyan:1991} They are particle-hole condensates in contrast to particle-particle condensates in a superconductor. Because there are no exchange  requirements between a particle and a hole, the orbital wave function cannot constrain the spin wave function. Of particular interest is the angular momentum  $\ell=2$: its singlet counterpart has been suggested to be the cause of the pseudogap in high-temperature superconductors.~\cite{Chakravarty:2001,Laughlin:2014a,Laughlin:2014b} Physically, it reflects staggered circulating charge currents in a two-dimensional square lattice. The triplet counterpart consists of circulating staggered spin currents but not charge currents. Recently an attempt~\cite{Fujimoto:2011} was made to relate it to the HO phase of URS to explain the in-plane anisotropic magnetic susceptibility $\chi[110] \neq \chi[\bar{1}10]$ observed in the magnetic torque measurements.~\cite{Okazaki:2011,Shibauchi:2012}
In Ref.~\citenum{Fujimoto:2011}, the triplet $d$-density wave is assumed to be formed on the diagonal planes ($x+y=$ constant), which breaks the $C_{4}$ rotational symmetry down to $C_{2}$. Therefore, the off-diagonal term of the spin susceptibility $\chi_{ab}$ arises below $T_{\textrm{HO}}$ and naturally explains the in-plane anisotropic susceptibility,
\begin{equation}
\chi[110] = \chi_{aa} + \chi_{ab} \neq \chi[\bar{1}10] =  \chi_{aa} - \chi_{ab}. 
\end{equation}
While this is an interesting idea, so far it has not been able to provide a mechanism for superconductivity, which must be related to the HO state.  

In addition to the in-plane anisotropic susceptibility, an Ising anisotropy of the $g$-factor has been observed in URS.~\cite{Altarawneh:2012,Ohkuni:1999} However, the Ising anisotropy does not lower the symmetry, while $C_{4}\to C_{2}$ is a broken symmetry.  In the presence of an external field, say $H=10$T, the Zeeman energy difference due to the anisotropic $g$-factor is
\begin{equation}
\Delta E_{\textrm{Zeeman}}=(g_{c}-g_{a}) \mu_{B} H=1.53 {\rm meV}
\end{equation} 
where $(g_{c}-g_{a}) \approx 2.65$ is the difference between the $g$-factor along the $c$ and $a$ axes,~\cite{Altarawneh:2012} and $\mu_{B}$ is the Bohr magneton. The magnitude of $\Delta E_{\textrm{Zeeman}}$ is an order magnitude smaller than the gap parameters of the st-DDW state estimated from the specific heat calculation (c.f. below). Therefore, the skyrmionic spin texture is unlikely to be affected by the anisotropic $g$-factor. 

In contrast, we consider a mixed st-DDW, which mixes the triplet and the singlet density waves in the $\ell=2$ particle-hole channel; see Refs.~\citenum{Hsu:2011,Hsu:2013} for details. Ref.~\citenum{Hsu:2013} introduced skyrmions as a spin texture in the st-DDW state. As has been analyzed in Ref.~\citenum{Hsu:2013}, the inclusion of the explicit spin-orbit coupling will not affect the skyrmionic texture in the st-DDW state even for the U atoms. The skyrmions were shown to have zero angular momentum and charge $2e$ bosons, so one could only predict  $s$-wave Bose-Einstein condensate. However, available experiments~\cite{Kasahara:2007,Yano:2008} show that the superconductivity is not $s$-wave, but chiral $d$-wave, breaking TRS. In this paper we propose a totally new unconventional pairing mechanism arising from the fractionalization of skyrmions into merons and antimerons. The mechanism resolves the paradox that skyrmions may have zero angular momentum, but the superconductivity can be a chiral $d$-wave condensate. We also predict a nonzero signal of the PKE at the onset of the superconductivity. None of these were contained in Ref.~\citenum{Hsu:2013}, nor could it have been simply guessed at. Moreover, the generalization to three dimensions to accommodate the quantum oscillation experiments,~\cite{Hassinger:2010,Nakashima:2003} as opposed to two dimensions in Ref.~\citenum{Hsu:2013}, is important.
 
The following is a summary of the experimental aspects of URS which the st-DDW state and the resulting superconductivity may account for:
\begin{itemize}
\item The nesting vector of the st-DDW, $\vec{Q}=\frac{2\pi}{c}\hat{z}$, gives rise to the unit cell doubling along the $c$ axis in agreement with the quantum oscillation measurements.~\cite{Hassinger:2010}

\item The broken $C_{4}$ rotational symmetry by the st-DDW on the diagonal planes results in the observed anisotropic susceptibility in the HO phase in the magnetic torque experiments.~\cite{Okazaki:2011,Shibauchi:2012}
 
\item The opening of the st-DDW gap causes an exponential behavior below $T_{\textrm{HO}}$ in the specific heat measurements.~\cite{Palstra:1985}

\item Because of the pairing mechanism discussed in this paper, the superconductivity arises only within the HO phase, but not the pressure-induced large-moment antiferromagetic phase.~\cite{Mydosh:2011}

\item The chiral $d$-wave superconducting order parameter has nodal structures consistent with the thermodynamic experiments.~\cite{Kasahara:2007,Yano:2008}

\item The broken TRS in the superconducting state is consistent with a nonzero signal of the PKE.~\cite{Kapitulnik}
\end{itemize}

The structure of this paper is as follows: in Sec.~\ref{Sec:topology}, we discuss the topology of the st-DDW state and the charge-2$e$ skyrmions. In Sec.~\ref{Sec:pairing}, we study the pairing interaction due to the merons and antimerons, and show that the superconductivity is chiral $d$-wave. In Sec.~\ref{Sec:discussion} we remark on our results and the experiments of $\mathrm{URu_{2}Si_{2}}$. In the Appendix, the calculation of the topological invariant and the charge of the skyrmions are provided.

\section{\label{Sec:topology}Nontrivial topology and charge-2$e$ skyrmions}

\begin{figure}
	\centering
	\includegraphics[width=0.45\linewidth]{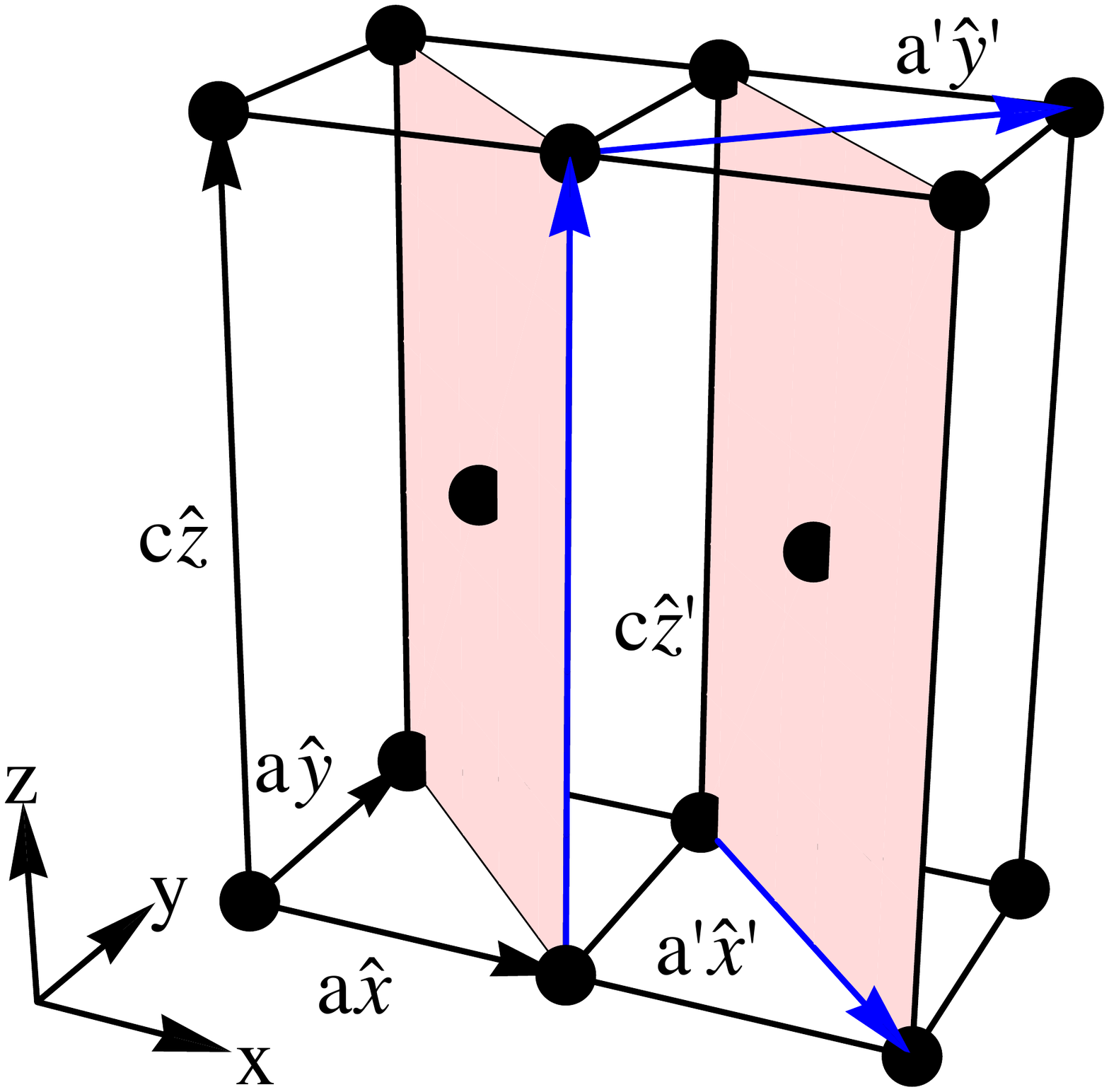}
	\includegraphics[width=0.3\linewidth]{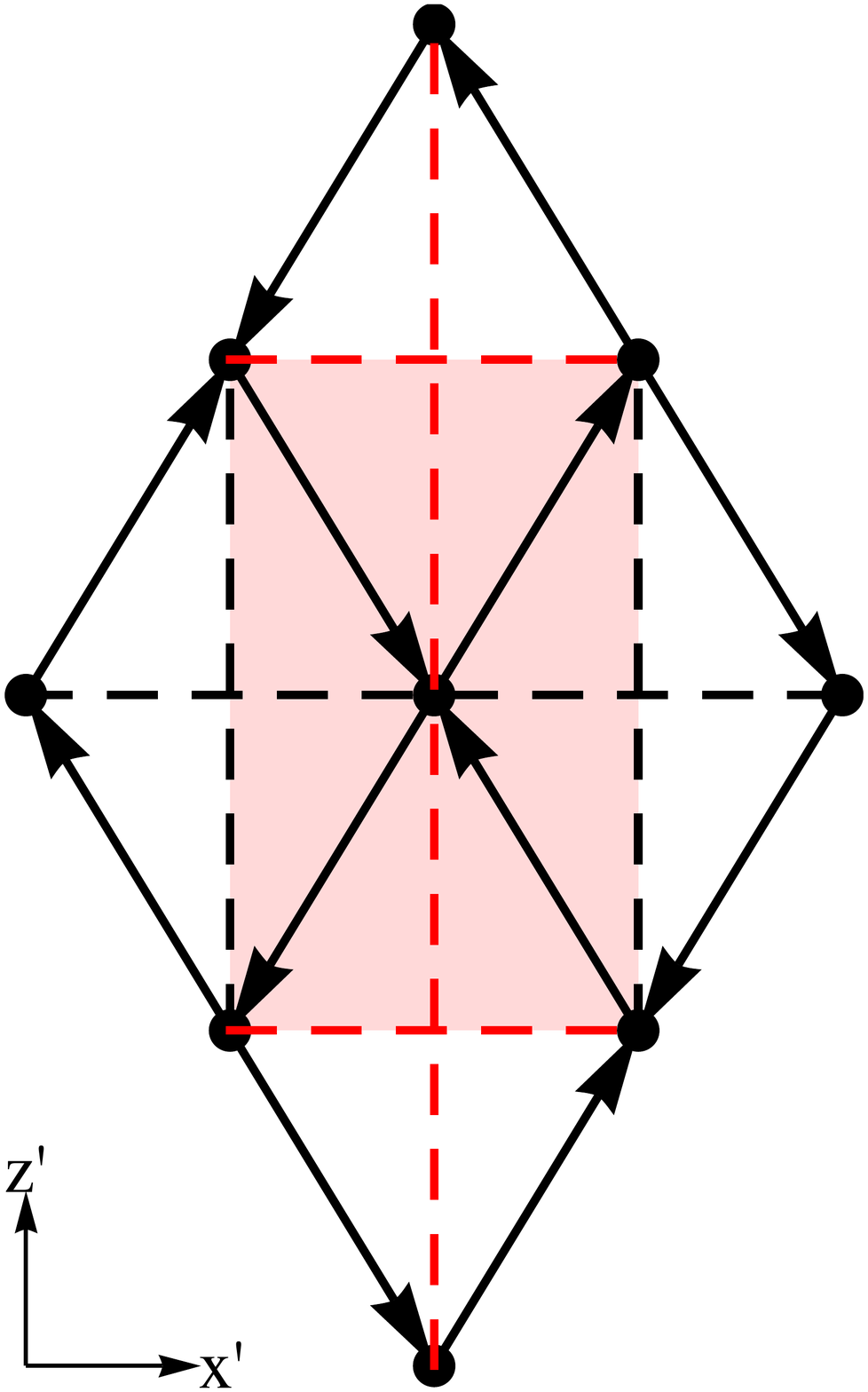} 
	\caption{Left: the crystal structure of the URS material. The modulated hopping and spin current patterns are on the diagonal planes ($x+y=$ constant), which is highlighted with the pink color. The black points indicate the positions of the U atoms. The black arrows indicate three primitive vectors, $a\hat{x}$, $a\hat{y}$ and $c\hat{z}$. The blue arrows indicate three vectors for the rotated coordinate: $a'\hat{x}'$, $a'\hat{y}'$ and $c\hat{z}'$. Here $\hat{x}'=(\hat{x}-\hat{y})/\sqrt{2}$, $\hat{y}'=(\hat{x}+\hat{y})/\sqrt{2}$, $\hat{z}'=\hat{z}$, and $a'=\sqrt{2}a$. The coordinate rotation is convenient to describe the st-DDW order, which is assumed to be formed on the diagonal planes and consistent with the magnetic torque measurement.~\cite{Okazaki:2011} 
Right: the spin current patterns and hopping modulations on the diagonal planes. 
The arrows indicate the directions of the circulating spin currents due to the triplet component of the mixed st-DDW order. The red and black dashed lines indicate different signs of the modulated hopping terms due to the singlet component of the mixed st-DDW order. The Ru and Si atoms are not shown for clarity.}
\label{Fig:Pattern}
\end{figure}

URS has a body-centered-tetragonal structure, and the order parameter and the band structure must be consistent with it; see FIG.~\ref{Fig:Pattern}. We consider the tight-binding model~\cite{Rau:2012} with the URS crystal structure and the st-DDW order on the diagonal planes, which leads to the observed anisotropic magnetic susceptibility. This model is merely a low-energy effective Hamiltonian, which is sufficient to illustrate our mechanism of superconductivity, but clearly cannot capture all aspects of URS:
\begin{eqnarray}
\mathcal{H}_{0} &=& \sum_{k,\sigma} \left( \epsilon_{k}^{(1)} c_{1\sigma, k}^{\dagger} c_{1\sigma, k} + \epsilon_{k}^{(2)} c_{2\sigma, k}^{\dagger} c_{2\sigma, k} \right) \nonumber \\
&& +\sum_{k} \left( C_{k} c_{1+, k}^{\dagger} c_{2+, k} + C_{k}^{*} c_{1-, k}^{\dagger} c_{2-, k} +\textrm{H.c.} \right) \nonumber \\
&& +\sum_{k} \left( D_{k} c_{1+, k}^{\dagger} c_{2-, k} - D_{k}^{*} c_{1-, k}^{\dagger} c_{2+, k} +\textrm{H.c.} \right),
\end{eqnarray} 
where $c_{\alpha\sigma, k}^{\dagger}$ is the creation operator of $5f$ electron with band index $\alpha=1,2$ and spin index $\sigma=\pm$, and the band structure is
\begin{eqnarray}
\epsilon_{k}^{(\alpha)} &\equiv& 8t \cos \left( \frac{k_{x} a}{2} \right) \cos \left( \frac{k_{y} a}{2} \right) \cos \left( \frac{k_{z} c}{2} \right)  \nonumber \\
&&
+ 2t'_{\alpha} \left[ \cos (k_{x}a) + \cos (k_{y}a) \right] \nonumber \\
&& + 4t''_{\alpha} \cos (k_{x}a) \cos (k_{y}a) -\mu + \textrm{sgn}(\alpha)\frac{\Delta_{12}}{2},
\end{eqnarray} 
where $\textrm{sgn}(\alpha)=\pm 1$ for band indices $\alpha=1,2$, respectively. Here $a$ and $c$ are the lattice constants.
$t,t'_{\alpha}$, and $t''_{\alpha}$ are the hopping amplitudes along the body diagonals, in-plane axes, and in-plane diagonals, respectively. $\Delta_{12}$ is the crystal-field splitting and $\mu$ is the chemical potential. Notice that $t,t'_{\alpha}$, and $t''_{\alpha}$ describe the hopping terms of the $5f$ electrons between the U atoms, and our conclusion holds as long as $t$ is nonzero (see below and the Appendix). As in Ref.~\citenum{Rau:2012}, we will take $C_{k}=0$. $D_{k}$ is related to the hybridization due to the Ru atoms, and has the form
\begin{eqnarray}
D_{k} &\equiv& 4t_{12} \left[ 
\sin \left( \frac{k_{x} a + k_{y} a}{2} \right) - i \sin \left( \frac{k_{x} a - k_{y} a}{2} \right)\right] \nonumber \\
&& \times \sin \left( \frac{k_{z}c}{2} \right)
\end{eqnarray}

Then the st-DDW order parameter is defined to be
\begin{equation}
\langle c_{\alpha \sigma,k+Q}^{\dagger} c_{\alpha \sigma',k} \rangle = \delta_{\sigma \sigma'}  \Delta_{k} +i (\vec{\sigma} \cdot \hat{N})_{\sigma \sigma'} W_{k},
\end{equation}
where $\sigma$ and $\sigma'$ are spin indices and the nesting vector is $\vec{Q}=\frac{2\pi}{c}\hat{z}$, consistent with the fact that quantum oscillation frequencies are hardly changed between HO and the large moment antiferromagnetic phase (LMAF);~\cite{Nakashima:2003,Hassinger:2010} $W_k$ and $\Delta_{k}$ are the form factors for the triplet and the singlet components of the density wave order:
\begin{eqnarray}
W_k  &=& W_0 \sin \left( \frac{k_{x} a - k_{y} a}{2} \right) \sin \left( \frac{k_{z} c}{2} \right), \\
\Delta_k &=& \frac{\Delta_0}{2} \left[ \cos \left( k_{x} a - k_{y} a   \right)  - \cos \left( k_{z} c \right) \right].
\end{eqnarray}
Judging from microscopic Hartree-Fock  calculations of Ref.~\onlinecite{Ikeda:1998} and Refs.~\onlinecite{Laughlin:2014a,Laughlin:2014b}, it is clear that no fine tuning is necessary to have such an   order parameter at the Hartree-Fock level. 
The above form factors are constrained by the nesting vector, $(0,0,\frac{2\pi}{c})$, in quantum oscillation experiments~~\cite{Nakashima:2003,Hassinger:2010} and the $C_{4}\to C_{2}$ rotational symmetry breaking.~\cite{Okazaki:2011} In contrast, the nesting vector adopted in Ref.~\onlinecite{Fujimoto:2011} is  $(0,0,\frac{\pi}{c})$.

We have checked that the inclusion of the hybridization, $t_{12}$, barely alters the ground-state energy. To be explicit, we found the change in the ground-state energy per lattice site is 
\begin{equation}
\Delta E_{\text{hyb}}\approx - \left( \frac{|t_{12}|^{2}}{W} \right) 
\left( \frac{W_{0}}{W} \right)^{2},
\end{equation}
where $W=8t$ is the bandwidth. $\Delta E_{\textrm{hyb}}$ is much smaller than $W_{0}$ and $\Delta_{0}$ because $(W_{0}/W)^{2}\approx 0.01$. Therefore, the inclusion of the hybridization will not affect our conclusion. For the calculation of the Berry phase crucial to our mechanism of superconductivity one only requires  the $t$-term in the band structure, hence the results are identical for the two bands; see Appendix~\ref{Topology}.

Furthermore, it has also been pointed out that the Fermi surface does not depend strongly on $t_{12}$,~\cite{Rau:2012} so $t_{12}$ will be neglected in the following discussion. 
In addition, the inclusion of the explicit spin-orbit coupling in the two-dimensional st-DDW model has been analyzed in Ref.~\citenum{Hsu:2013}, and it has been shown that the change in the ground-state energy per lattice site is \begin{equation}
\Delta E_{\textrm{SO}}\approx [(\hat{N}\cdot \hat{z})^{2}-1] \left( \frac{\Lambda_{0}^{2}}{W} \right) \left( \frac{W_{0}}{W} \right)^{2}
\left[ 1+ O(\frac{W_{0}^{2}}{W^{2}}) \right].
\end{equation}
Here $\Lambda_{0}$ is the strength of the spin-orbit coupling, given by
\begin{equation}
 \mathcal{H}_{\textrm{SO}}=\sum_{k} c^{\dagger}_{k\alpha}\vec{\Lambda}(k)\cdot \vec{\sigma}_{\alpha\beta} c_{k\beta}, 
\end{equation}
where $\vec{\Lambda}(k) = (\Lambda_{0}/\sqrt{2})[\hat{x} \sin k_{y}-\hat{y} \sin k_{x} ]$. As large as the spin-orbit coupling may be for U atoms, $\Delta E_{\textrm{SO}}$ is still a small energy scale because of the small factor $(W_{0}/W)^{2} \approx 0.01$. The analysis in Ref.~\citenum{Hsu:2013} is based on the second-order expansion in $(W_{0}/W)$, which is justified because $W_{0}$ is one order smaller than $W$. 
Therefore, the charge-2$e$ skyrmionic texture that we invoke below is unlikely to be affected by the explicit spin-orbit coupling, which will be neglected in the discussion. 
Notice that the order parameter itself cannot be factorized into spin and orbital parts, so it requires spin-orbit interaction to be realized. In other words, the spin-orbit interaction is present in the model even though we do not include it in the Hamiltonian explicitly.   

At the mean-field level, we can choose the $\hat{N}$ vector to be uniform and perpendicular to the diagonal planes ($x+y=$ constant). The real space picture of the order parameter form factors are shown in FIG.~\ref{Fig:Pattern}.
Note that this is different from Ref.~\citenum{Fujimoto:2011}, where there are two copies of spin current patterns on the diagonal planes and the singlet component is missing.~\cite{Hsu:2013} The spin currents are unaffected by the singlet component, as that only produces modulations of the bare kinetic energy. Therefore, the $C_{4}$ rotational symmetry is broken by the st-DDW order, which results in anisotropic susceptibility as in Ref.~\citenum{Fujimoto:2011}. However, the singlet component $\Delta_{0}$ has an important consequence on the basic nature of the HO state. One of the experimental signatures of the HO state is a jump in the specific heat $\frac{\Delta C}{T} \approx 270$ mJ/mol-K$^2$ at $T_{\textrm{HO}}$, followed by an exponential drop below $T_{\textrm{HO}}$, which can be fitted with a gap of $\approx$ 11 meV.~\cite{Palstra:1985} 
The specific heat for the mixed st-DDW state also exhibits a similar  exponential behavior when we consider the fully gapped $k_{y'}a'=\pi$ plane; see FIG.~\ref{Fig:Pattern} for the rotation of the coordinate axes. 
This implies that the specific heat reflects primarily the quasi-two-dimensional part of the spectrum, which justifies that it is a good approximation to consider the system as a collection of quasi-two-dimensional diagonal planes, with low carrier concentration. With the gap parameters of $W_{0}= 14$ meV and $\Delta_{0}= 13$ meV, we obtained the exponential drop $C(T) \propto e^{-\frac{\Delta}{T}}$, which is consistent with experiments, except for lower temperature linear behavior due to the fact that the three-dimensional Fermi surface is only partially gapped. Importantly, our basic mechanism of superconductivity depends on the quantized spin Hall effect on the diagonal planes, which is absent if $\Delta_{0}=0$  (cf. below).

The mean-field Hamiltonian with the mixed st-DDW order is
\begin{equation}
\mathcal{H}_{\textrm{st-DDW}}=\sum_{k} \Psi_{k}^{\dagger} A_k \Psi_k,
\end{equation}
where the summation is over the reduced Brillouin zone (RBZ). The spinor, $ \Psi_{k}^{\dagger}$, is defined in terms of the fermion operators $(c_{k,\uparrow}^{\dagger}, c_{k+Q,\uparrow}^{\dagger}, c_{k,\downarrow}^{\dagger}, c_{k+Q,\downarrow}^{\dagger}) $ and the matrix $A_k$ is
\begin{equation}
 A_k =\left(
  \begin{array}{cccc}
   \epsilon_{k}-\mu  & \Delta_k+iW_k      & 0                 & 0             \\
   \Delta_k-iW_k     & \epsilon_{k+Q}-\mu & 0                 & 0             \\
   0                 & 0                  & \epsilon_{k}-\mu  & \Delta_k-iW_k \\
   0                 & 0                  & \Delta_k+iW_k     & \epsilon_{k+Q}-\mu
  \end{array}
\right),
\end{equation} 
where 
\begin{eqnarray}
\epsilon_k &\equiv& 8t \cos \left( \frac{k_{x} a}{2} \right) \cos \left( \frac{k_{y} a}{2} \right) \cos \left( \frac{k_{z} c}{2} \right) \nonumber \\
&& + 2t' \left[ \cos (k_{x}a) + \cos (k_{y}a) \right] \nonumber \\
&& + 4t'' \cos (k_{x}a) \cos (k_{y}a),
\end{eqnarray}
where $t,t'$, and $t''$ are the hopping amplitudes along the body diagonals, in-plane axes, and in-plane diagonals, respectively.
Although a two-band tight-binding model is considered above,~\cite{Rau:2012} the Berry curvature for these two bands are the same because the $t'$, $t''$ and $\Delta_{12}$ terms commute with the Hamiltonian. Therefore, for simplicity the band index $\alpha$ and the crystal-field splitting $\Delta_{12}$ have been dropped.

The eigenvalues of the Hamiltonian are
\begin{equation}
 \lambda_{k,\pm}= \epsilon_{2k}-\mu \pm E_{k},
\end{equation}
where $E_{k}=\sqrt{ \epsilon_{1k}^2 + W_k^2 + \Delta_k^2}$, the $+$($-$) sign indicates the upper (lower) band, and
\begin{eqnarray}
\epsilon_{1k} &\equiv& {\frac {\epsilon_k - \epsilon_{k+Q}}{2}} \nonumber \\
&=& 8t \cos \left( \frac{k_{x} a}{2} \right) \cos \left( \frac{k_{y} a}{2} \right) \cos \left( \frac{k_{z} c}{2} \right), \\
\epsilon_{2k} &\equiv& {\frac {\epsilon_k + \epsilon_{k+Q}}{2}} \nonumber \\
&=& 2t' \left[ \cos (k_{x}a) + \cos (k_{y}a) \right] + 4t'' \cos (k_{x}a) \cos (k_{y}a) \nonumber \\
\end{eqnarray}

The eigenvalues and eigenvectors of the mean-field Hamiltonian can be used 
to compute the Berry curvature, $\vec{\Omega}_{\sigma,\pm}$, for the upper and the lower bands ($\pm$),
\begin{equation}
\vec{\Omega}_{\sigma,\pm} \equiv \vec{\bigtriangledown}_k \times \langle \Phi_{\sigma,\pm}(k)| i \vec{\bigtriangledown}_k | \Phi_{\sigma,\pm}(k) \rangle,
\end{equation}
where $|\Phi_{\sigma,\pm}(k) \rangle$ are the corresponding eigenstates. The Berry curvature is necessary for the computation of the physical charge and flux carried by the skyrmions.  
Since the mixed st-DDW order is on the diagonal planes, the nonzero contribution to the Berry phase arises from the component of the Berry curvature perpendicular to the diagonal planes. The result does not depend on the details of the band parameters as long as $t,W_{0}$, and $\Delta_{0}$ are all nonzero (see the appendix). In other words, we need a mixing of the triplet and the singlet density wave orders in order to have nontrivial topology. 
As in the previous two-dimensional model,~\cite{Hsu:2011} the total Chern number is zero, but the spin Chern number is nonzero. Therefore, the topology of the system is nontrivial, and there will be a quantized spin Hall conductance on the $x'z'$ planes, $\sigma_{x'z'}^{\textrm{spin}} = {\frac {e}{2\pi}}$. Because of this the charge current corresponds to a physical charge (see the Appendix).
Then, the skyrmionic spin texture can be constructed on the $x'z'$ plane, and one can find that the skyrmions in the system carry physical flux $4\pi$, as in Refs.~\citenum{Grover:2008,Hsu:2013}. As a skyrmion is adiabatically threaded through the system, a net charge of $-2e$ is displaced to the boundary at infinity; by charge neutrality of the total system, the skyrmion should have physical charge $2e$ and flux $4\pi$.

\section{\label{Sec:pairing}chiral $d$-wave pairing}

\begin{figure}
	\centering
	\includegraphics[width=0.75\linewidth]{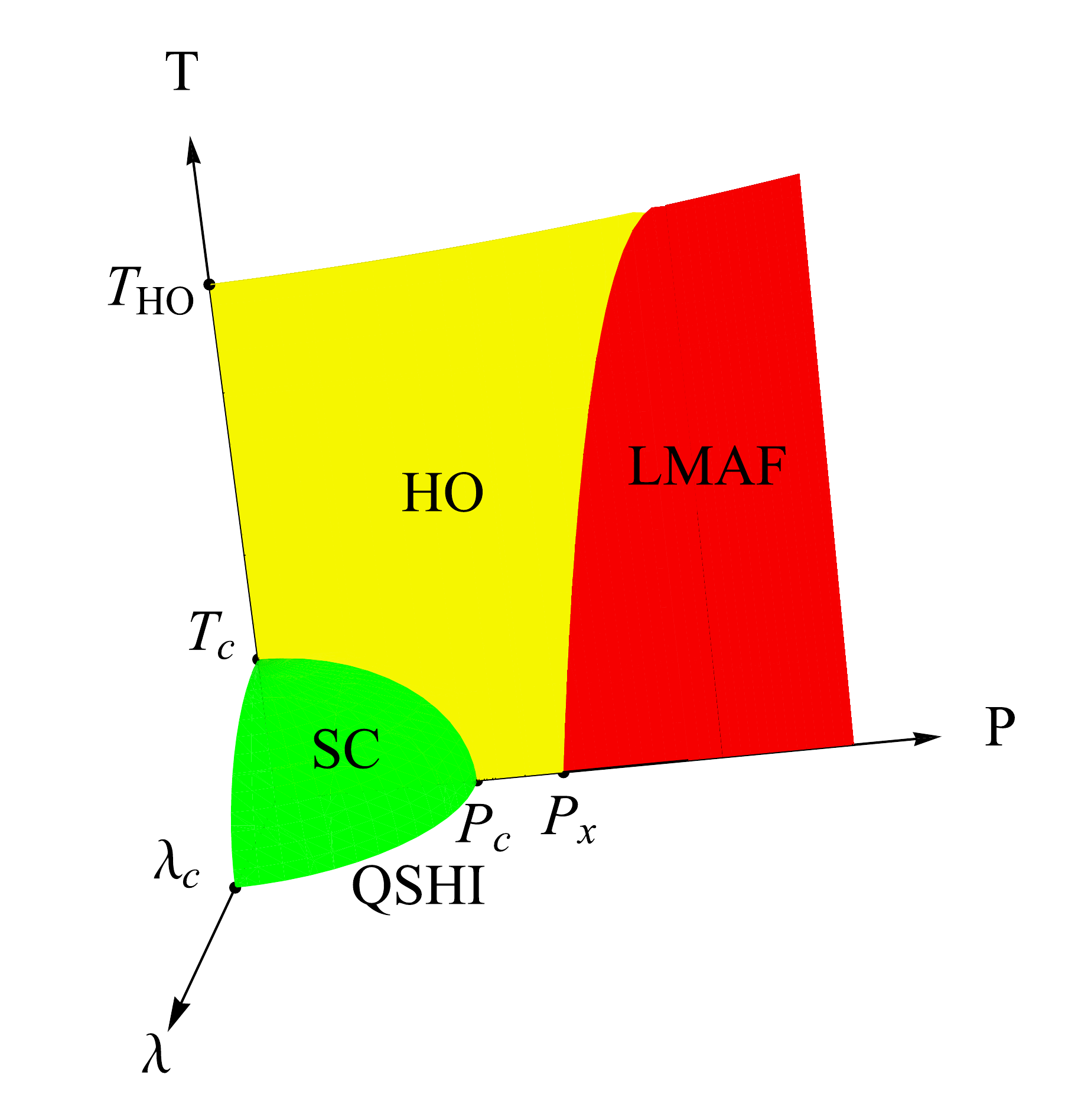} 
	\caption{The proposed phase diagram with the quantum parameter ($\lambda$), pressure ($P$), and temperature ($T$) axes. Here $\lambda$ is a tuning parameter such that  $W_{0}(\lambda<\lambda_{c})=0$ and $W_{0}(\lambda>\lambda_{c})\neq 0$. $\lambda_{c}$ is a deconfined quantum critical point between the QSHI and superconductivity as $T=P=0$. $T_{\textrm{HO}}$ and $T_{c}$ are the HO and superconducting transition temperatures as $P=\lambda=0$, respectively. Along the $P$ axis, $P_{c}$ indicates the phase transition between the HO and superconducting states, while $P_{x}$ indicates the phase transition between the HO and LMAF states. In some literatures $P_{c}$ coincides with $P_{x}$, which does not affect our main conclusion. In addition, there should be phase boundaries in the $\lambda$-$P$ and $\lambda$-$T$ planes, which are not the main purpose of this work.}
	\label{fig:PD}
\end{figure}

A useful way to proceed is to sketch a proposed phase diagram in which we introduce a quantum parameter $\lambda$ in addition to the  parameters pressure, $P$, and temperature $T$, as shown in FIG.~\ref{fig:PD}.
$\lambda$ controls  $W_{0}(\lambda)$ such that $W_{0}(\lambda<\lambda_{c})=0$ and $W_{0}(\lambda>\lambda_{c})\neq 0$. Note that an isolated meron costs logarithmically infinite energy for $\lambda > \lambda_{c}$, and hence merons and antimerons appear as bound pairs in skyrmions. The length scale of the confinement potential grows when approaching the deconfined quantum critical point, where it diverges. Therefore, the  skyrmions fractionalize into merons and antimerons, because there is no confinement at that point. We must emphasize that the hedgehog configurations are assumed to be suppressed because the particle-hole excitations are of much higher energy.~\cite{Grover:2008} Therefore the skyrmion number is conserved in the two-dimensional diagonal planes. 
The state at $T=0,P=0$ is connected, as is the entire superconducting state, by continuity from the second order phase transition between the superconductivity and quantum spin Hall insulator (QSHI); $\lambda_{c}$ is a deconfined quantum critical point.~\cite{Grover:2008} The suppression  of hedgehog configurations is crucial to the existence of the deconfined quantum critical point.
This critical point is described by the field theory of merons and antimerons, fractional particles that emerge at $\lambda_{c}$, but are not present in either side of it. $\lambda_{c}$ can be computed from a suitable microscopic Hamiltonian; for instance, $\lambda$ may be a function of the on-site Coulomb interaction $U$, the nearest neighbor direct interaction $V$, and the exchange interaction $J$ in the extended Hubbard model.~\cite{Ikeda:1998}

Since merons have topological charge they should have zero overlap with band fermions and therefore cannot be expressed in terms of local band fermonic operators. This is not unprecedented. Recall that creation/annihilation of Laughlin quasiparticles in the fractional quantum Hall effect cannot be expressed as any local function of the band fermions.

\begin{figure}
	\centering
	\includegraphics[width=\linewidth]{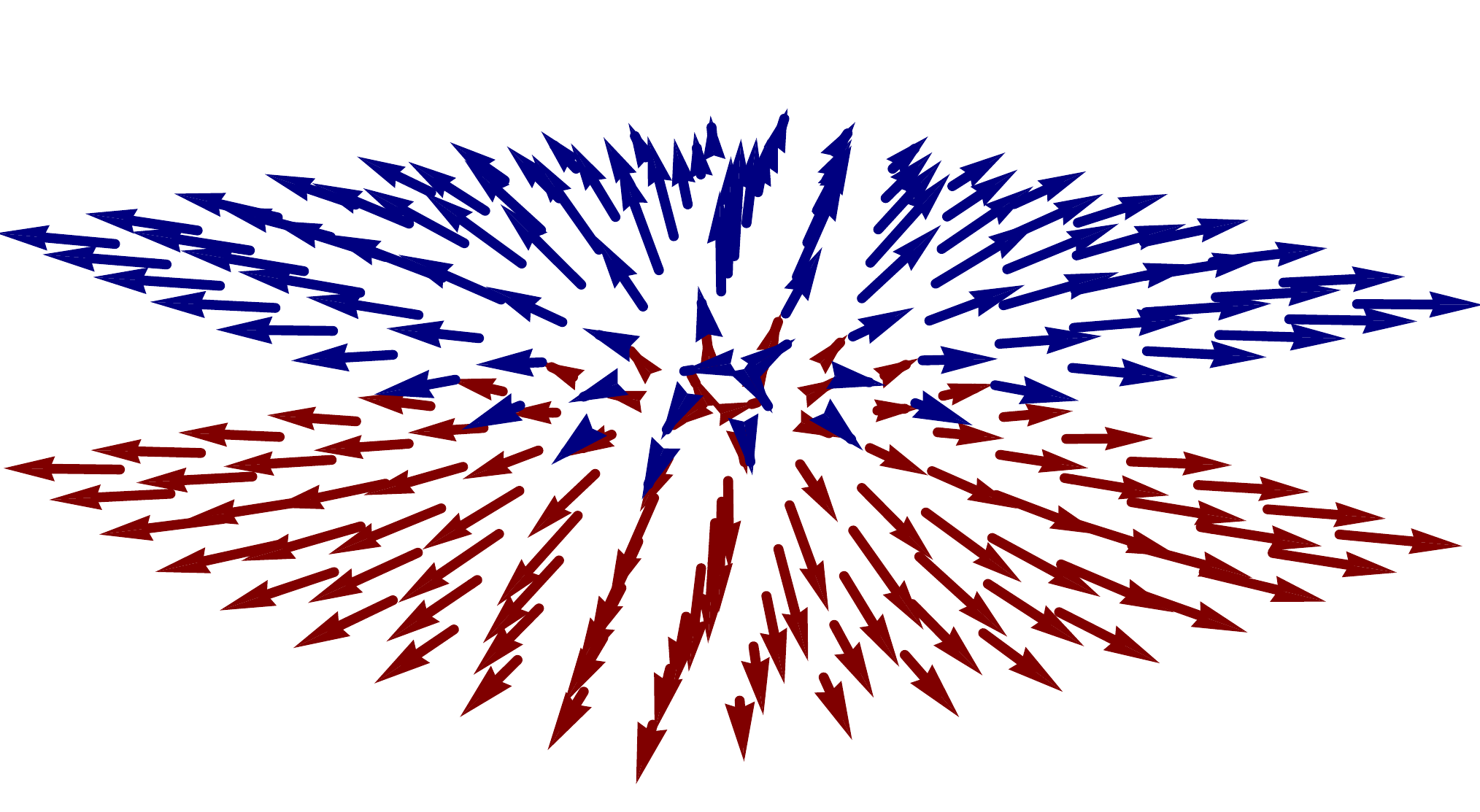} 
	\caption{The merons $\psi_{+,\sigma}^{\dagger}(\vec{r})$ and $\psi_{-,\sigma}^{\dagger}(\vec{r})$. $\psi_{+,\sigma}^{\dagger}(\vec{r})$ creates a meron with $\hat{N}(r\rightarrow 0)=\hat{y'}$ and $\hat{N}(r\rightarrow \infty)=\frac{(x',0,z')}{r}$; $\psi_{-,\sigma}^{\dagger}(\vec{r})$ creates a meron with $\hat{N}(r\rightarrow 0)=-\hat{y}'$ and $\hat{N}(r\rightarrow \infty)=\frac{(x',0,z')}{r}$. Each meron above is half a skyrmion. A composite of a meron $\psi_{+,\sigma}^{\dagger}(\vec{r})$ and an antimeron $\psi_{-,\sigma'}(\vec{r})$ makes one skyrmion.~\cite{Senthil:2004}}
	\label{Fig:composite}
\end{figure}

\begin{figure}
	\centering
	\includegraphics[width=0.25\linewidth]{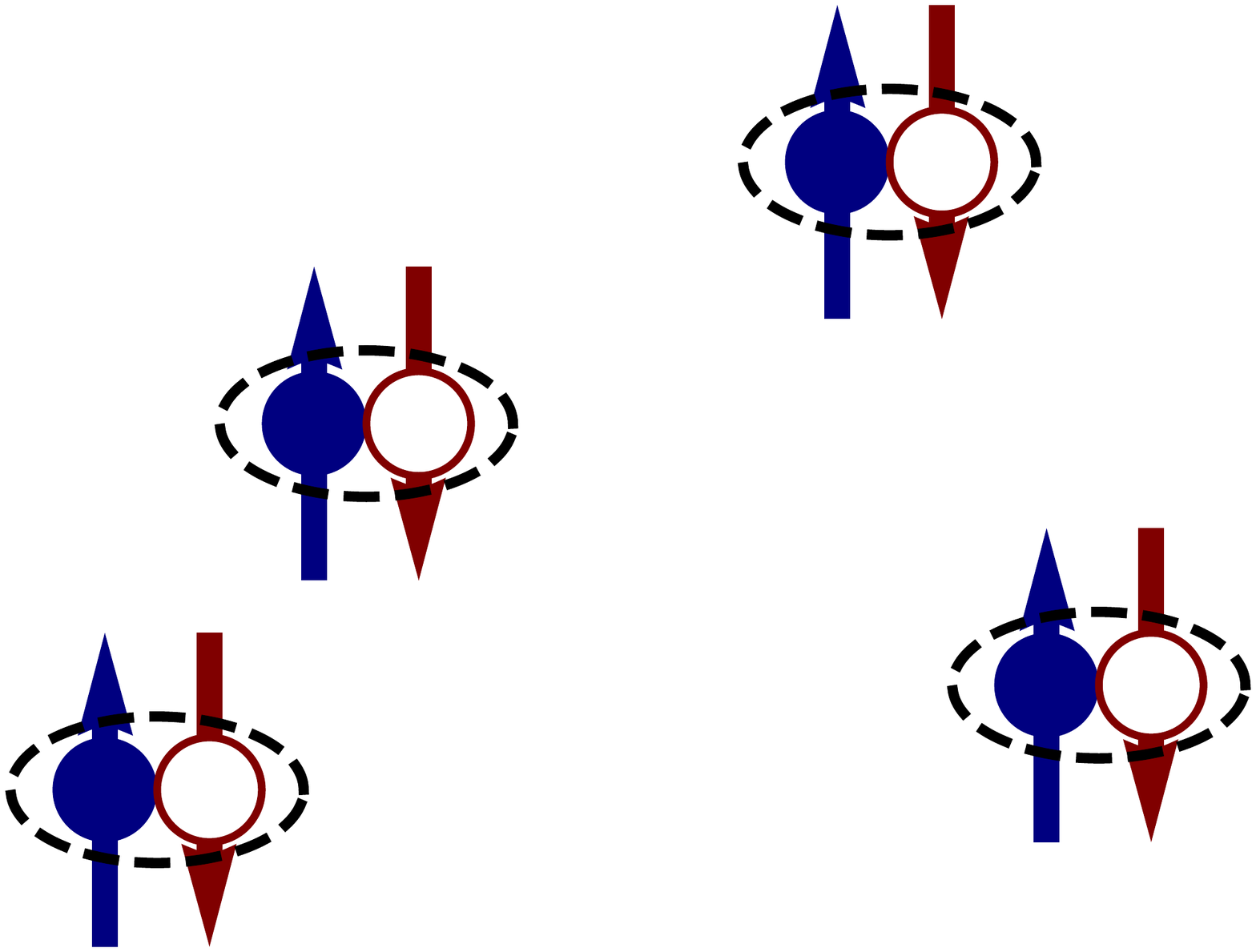}\hspace{0.1\linewidth} 
	\includegraphics[width=0.25\linewidth]{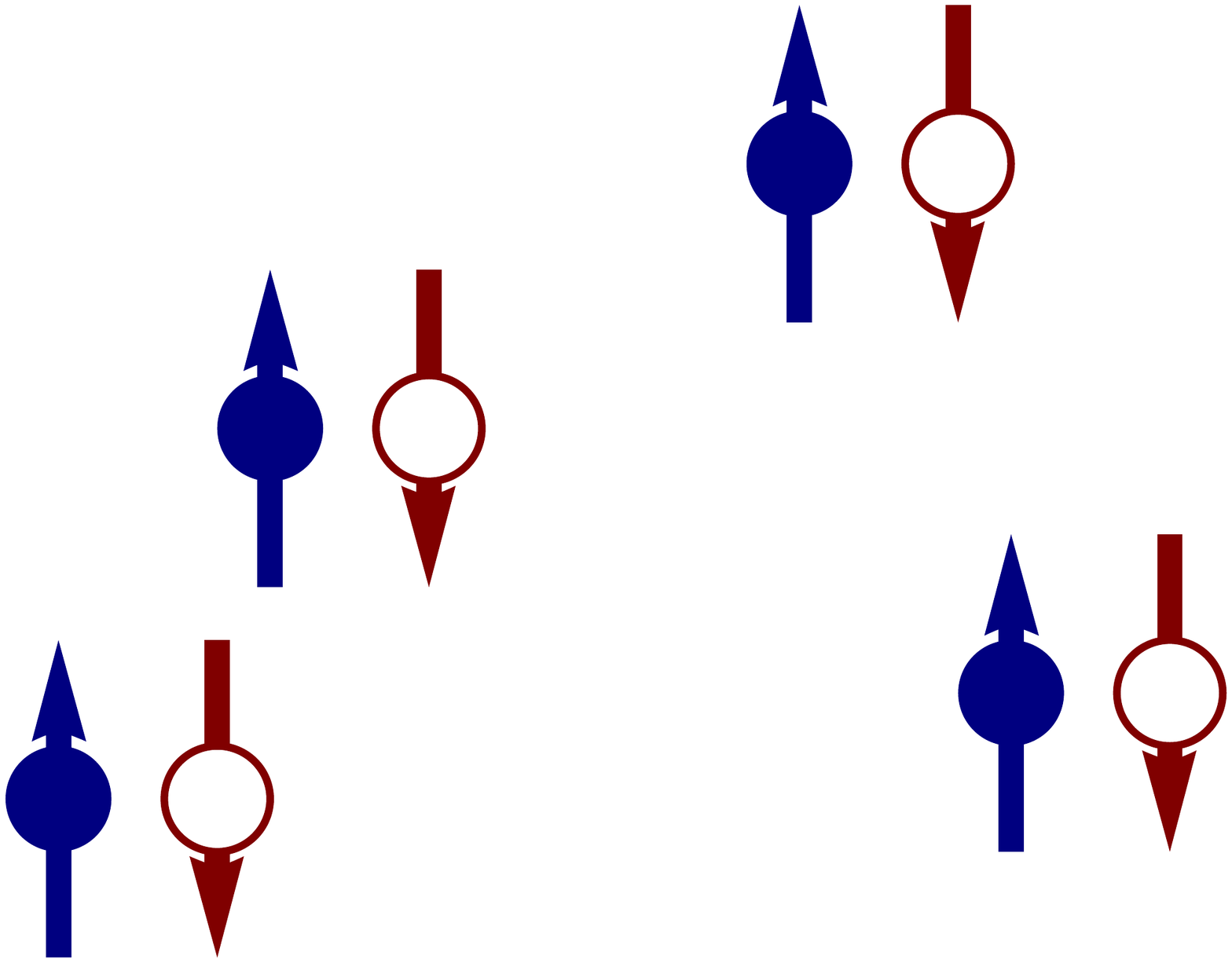}\hspace{0.1\linewidth} 
	\includegraphics[width=0.25\linewidth]{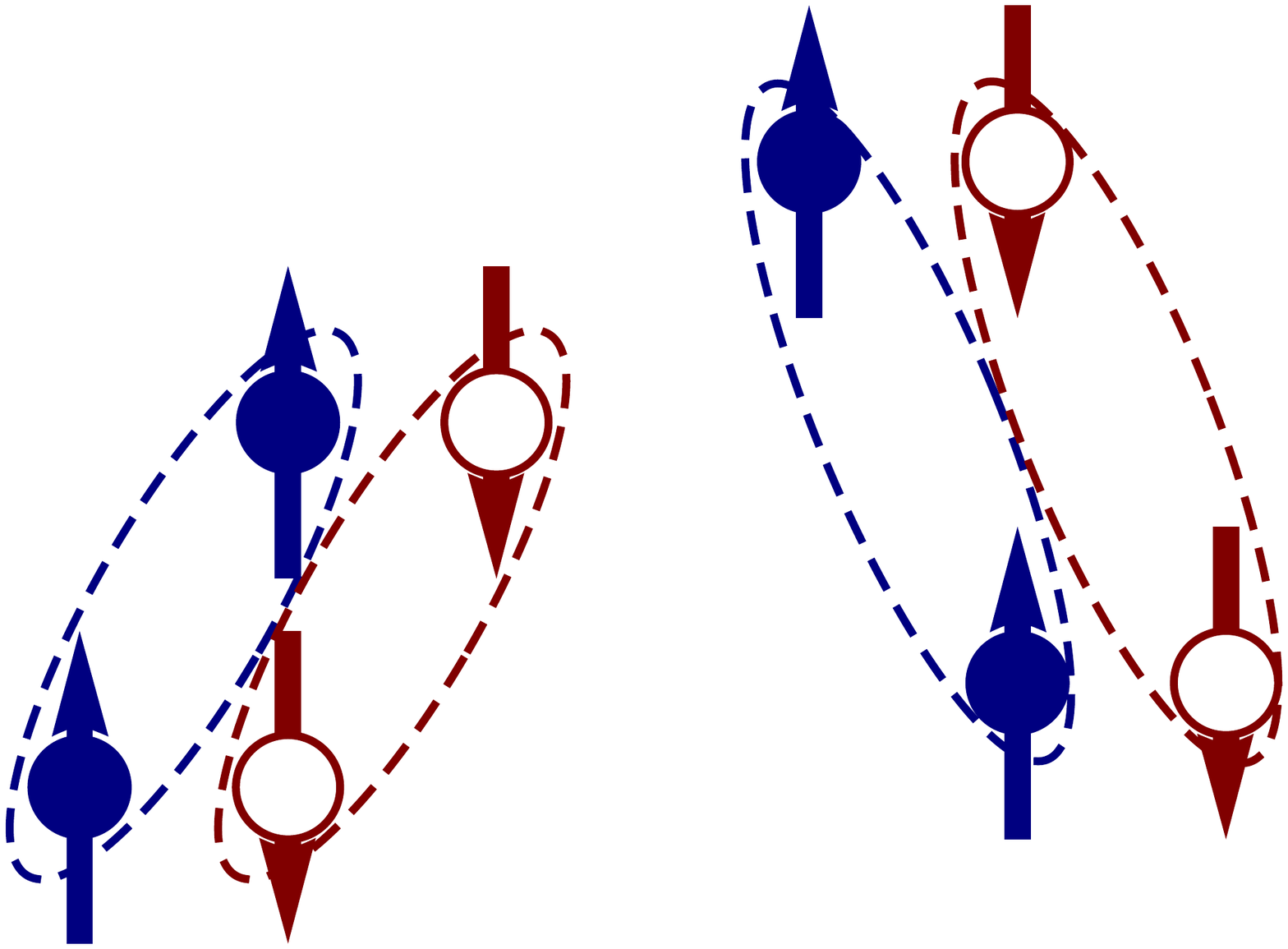}
	\caption{The deconfinement and pairing of the merons and antimerons. 
The up (down) arrows indicate a flux of $2\pi(-2\pi)$. The solid (open) circles indicate merons (antimerons). The dashed lines indicate the confinement and pairing. The colors are associated with the meron texture in FIG.~\ref{Fig:composite}. The spins of the merons are not shown for simplicity.
Left: in the HO phase, merons and antimerons are confined in skyrmions. Middle: at the critical point, the merons and antimerons are deconfined. Right: in the superconducting phase, the merons and antimerons are separately confined again into Cooper pairs.}
	\label{Fig:deconfine}
\end{figure}

A skyrmion is a composite of a meron with a flux of $2\pi$ and charge $e$, and an antimeron with a flux of $-2\pi$ and charge $-e$, as shown in FIG.~\ref{Fig:composite}. In the HO phase, the fractional particles are confined in skyrmions while in the superconducting phase they are bound into Cooper pairs (see FIG.~\ref{Fig:deconfine}).
Let $\psi_{s,\sigma}^{\dagger}(\vec{r})$ be the creation operator of a meron at $\vec{r}$, where $s=\pm$ labels the flux of $\pm 2\pi$ and the (pseudo)spin index $\sigma=\uparrow$ or $\downarrow$ for up or down spin, respectively. Pairing of $\langle \psi_{s,\sigma}^{\dagger}(\vec{r}) \psi_{s',\sigma'}^{\dagger}(\vec{r}')\rangle$ thus results in a charge $2e$ superconductivity for $s=s'$. Motivated by experiments,~\cite{Kasahara:2007} we will be interested in the (pseudo)spin-singlet pairing,~\cite{Sigrist:1991} so we will set $\sigma=-\sigma'$ . Here we assume that the length scale of a meron is much smaller than the distance between the merons, so we can treat them as point particles. 

The interaction Hamiltonian can be described by the coupling between the charge current $\vec{j}(\vec{r})$ and the gauge field $\vec{A}(\vec{r})$, which is associated with the flux of the merons and antimerons,
$
{\cal H}_{\textrm{int}} 
= \int d^2 r \vec{j}(\vec{r}) \cdot \vec{A}(\vec{r}) .
$
Notice that we will set $\hbar = c = 1$ for simplicity.
With the continuity equation for the charge and current densities of the merons and antimerons, and assuming that the kinetic energy of merons is $\frac{k^2}{2m}$ with the effective mass of a meron $m$, we may write the interaction in terms of the $\psi_{s,\sigma}(\vec{r})$ operators.
This immediately leads to the (pseudo)spin-singlet pairing Hamiltonian in the momentum space
\begin{eqnarray}
{\cal H}_{\textrm{int}} 
&=& \sum_{\vec{k},\vec{k}'} \sum_{s} V_{k'k}
\psi_{k',s,\uparrow}^{\dagger} \psi_{-k',s,\downarrow}^{\dagger}  \psi_{-k,s,\downarrow} \psi_{k,s,\uparrow}, 
\label{Eq:interaction2} 
\end{eqnarray}
where $V_{k'k} \equiv \frac{4 \pi i}{m} \frac{(\vec{k} \times \vec{k}')_{y'}}{|\vec{k} - \vec{k}'|^2}$. ${\cal H}_{\textrm{int}}$ clearly  breaks TRS, so we expect the superconducting order to break TRS as well.

The interaction is similar to the one discussed in the half-filled Landau level problem~\cite{Greiter:1992} as well as one proposed in the context of the hole-doped cuprates,~\cite{Morinari:2006} though there are significant differences as well. In Ref.~\citenum{Greiter:1992}, the flux attached to a particle is $\pi \epsilon$ instead of $2\pi$ so that fractional statistics could be studied by varying $\epsilon$. The particles considered there were spinless fermions, so an odd-parity pairing state was obtained. Furthermore, the interaction (\ref{Eq:interaction2}) is different from the one in Ref.~\citenum{Morinari:2006} because we express 
$
\rho(\vec{r}) = \sum_{s,\sigma} (se) \psi_{s,\sigma}^{\dagger}(\vec{r}) \psi_{s,\sigma}(\vec{r})
$ and 
$
\left( \vec{\triangledown} \times \vec{A}(\vec{r}) \right)_{y'}= \frac{2\pi}{e} \sum_{s,\sigma} s \psi_{s,\sigma}^{\dagger}(\vec{r}) \psi_{s,\sigma}(\vec{r})
$ differently. In Ref.~\citenum{Morinari:2006}, the resulting interaction depends on the sign of $s$, so the $s=\pm$ part leads to a $(d_{x^2-y^2} \mp id_{xy})$ superconductivity, respectively. As a result, the addition of these two components gives a $d_{x^2-y^2}$ superconductivity in cuprates, but {\em not} a chiral state. 
 
In Eq.~(\ref{Eq:interaction2}) we can see that the $s=+$ and $s=-$ parts in our case are two independent copies. With the kinetic energy term, the total Hamiltonian is
\begin{equation}
{\cal H}_{\textrm{total}} 
= \sum_{\vec{k},s, \sigma} \xi_k  \psi_{k,s,\sigma}^{\dagger}  \psi_{k,s,\sigma}  
+ {\cal H}_{\textrm{int}},
\end{equation}
where $\xi_{k} = \frac{k^2}{2m}-\mu$ with chemical potential $\mu$.
Defining the gap $\Delta_k^{sc} \equiv - \sum_{k'} V_{kk'} \langle \psi_{-k',\downarrow} \psi_{k',\uparrow} \rangle$, the BCS  gap equation at temperature $T$ is\begin{eqnarray}
\Delta_k^{sc}(T) &=& - \sum_{k'} V_{kk'} 
\frac{\Delta_{k'}^{sc}(T)}{2\sqrt{\xi_{k'}^2 + |\Delta_{k'}^{sc}(T)|^2 }} \nonumber\\
&& \hspace{0.3in} \times \tanh \left(
\frac{1}{2k_{B} T} \sqrt{\xi_{k'}^2 + |\Delta_{k'}^{sc}(T)|^2 } \right).
\end{eqnarray}
This  equation at $T=0$ has been analyzed in Refs.~\citenum{Greiter:1992, Morinari:2006}. For $\ell$ wave pairing, the solution will be
$
\Delta_k^{sc}(T) = |\Delta_k^{(\ell)}(T)| e^{-i \ell \phi_k},
$
where $\phi_{k}$ denotes the direction of the wave vector. The magnitude of the gap can be written as (see, Appendix~\ref{Appendix:D})
\begin{eqnarray}
|\Delta_k^{(\ell)}(T)| &=& \left\{
\begin{array}{l} 
\Delta_{F}^{(\ell)}(T) \left(\frac{k}{k_{F}}\right)^{\ell}, \textrm{~~~for~} k\le k_{F} \\
\Delta_{F}^{(\ell)}(T) \left(\frac{k_{F}}{k}\right)^{\ell}, \textrm{~~~for~} k\ge k_{F}
\end{array} \right.
\end{eqnarray}
with the Fermi wave vector $k_{F}$ and the temperature-dependent gap $\Delta_{F}^{(\ell)}(T)=|\Delta_{k=k_{F}}^{(\ell)}(T)|$.
The gap equation is  solved numerically, and nonzero solutions for $\Delta_{F}^{(\ell)}(T)$ is found for any $\ell \neq 0$; there is no solution for $\ell=0$. Since we are interested in the (pseudo)spin singlet pairing,~\cite{Kasahara:2007} we will focus on the even angular momentum channels in the following discussion.

\begin{figure}[htb]
\centering
\includegraphics[width=\linewidth]{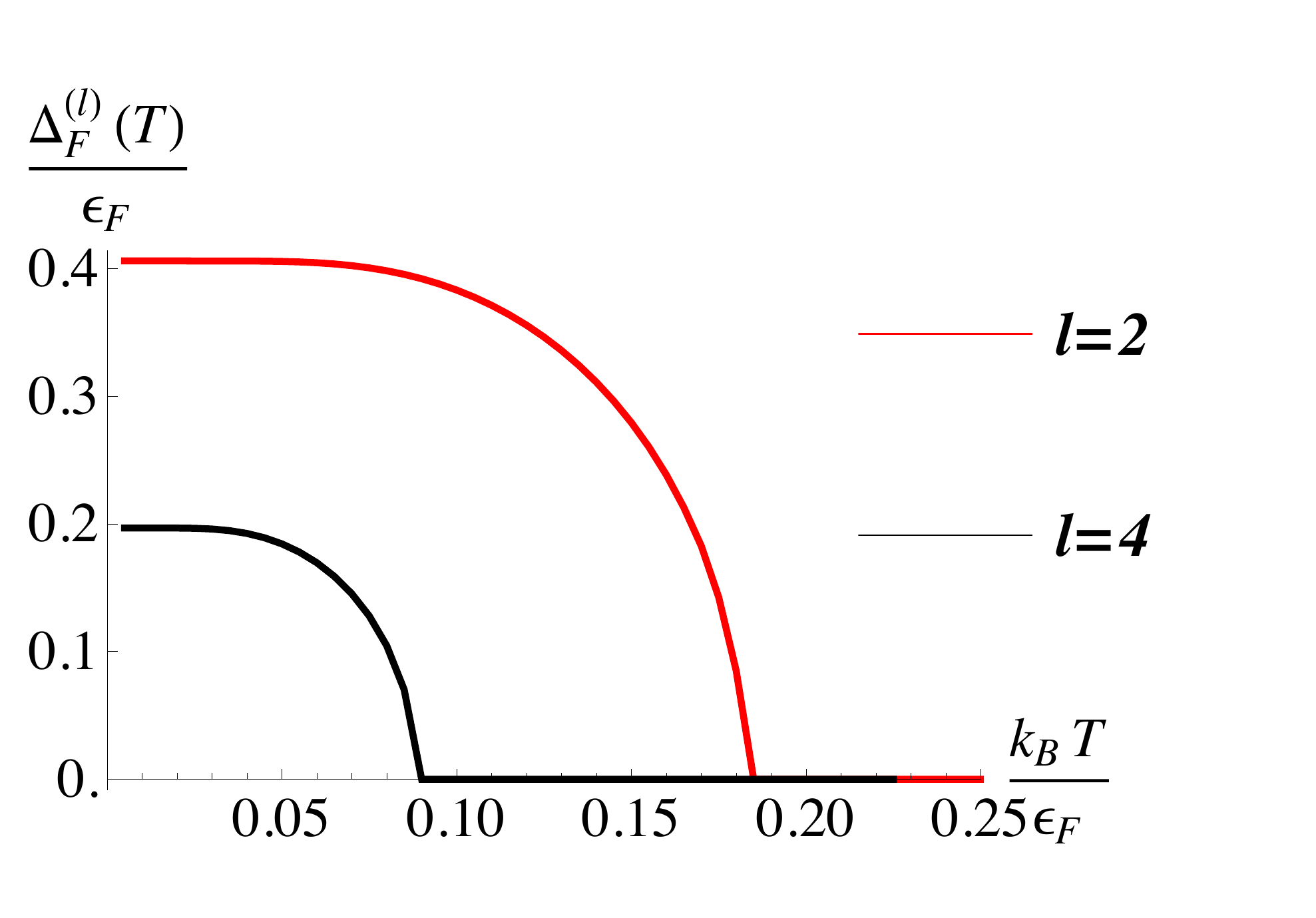}
\caption{Temperature dependence of the pairing gaps for the angular momentum channels $\ell=2,4$. 
}
\label{Fig:T-dep}
\end{figure}

The temperature dependence of the gap $\Delta_{F}^{(\ell)}(T)$ for $\ell=2,4$ is shown in FIG.~\ref{Fig:T-dep}. The zero temperature gap and superconducting transition temperature are listed in Table~\ref{Tab:gap}. Within a numerical prefactor the gap at $T=0$  is proportional to the Fermi energy $\epsilon_{F}$ of merons because there are no other available scales. 

From Table~\ref{Tab:gap} one can see that both the zero temperature gap and transition temperature for the $\ell=2$ channel are larger than the $\ell=4$ channel, so the dominant channel will be the chiral $d$-wave pairing. Clearly $\ell=4$ is a subdominant gap, which a thermodynamic measurement like specific heat will not be able to detect. The ratio $2\Delta_{F}^{(\ell=2)}(T=0)/(k_B T_{c}) \sim 4.390$ is comparable to the experimental value 4.9 from the point contact spectroscopy.~\cite{Morales:2009} Notice that both the theoretical and experimental values are larger than the BCS value $\sim 3.52$. Hence the superconducting gap in the continuum limit will be
\begin{equation}
\Delta_{k}^{sc}\propto k^2 e^{-2i \phi_{k}} = \left( k_{x'}^2 - k_{z'}^2 \right) - 2i  k_{x'}k_{z'} 
\label{Eq:d-gap}
\end{equation}
in the rotated coordinate. In the lattice, the gap function may be matched to
\begin{eqnarray}
\Delta_{k}^{sc} &\propto& \left[ \cos \left( k_{x'}a' \right)- \cos \left( k_{z'} c \right) \right]  \nonumber \\
&& + 4i \sin \left( \frac{k_{x'}a'}{2} \right) \sin \left( \frac{k_{z'}c}{2} \right).
\label{Eq:d-gap1}
\end{eqnarray}
Alternatively, another possible choice for the pairing function could be
\begin{eqnarray}
\Delta_{k}^{sc} &\propto& \cos \left( k_{y'} a'\right)
\left\{\left[ \cos \left( k_{x'}a' \right)- \cos \left( k_{z'} c \right) \right] \right.  \nonumber \\
&& \hspace{0.6in} \left.
  + 4i \sin \left( \frac{k_{x'}a'}{2} \right) \sin \left( \frac{k_{z'}c}{2} \right)\right\}.
\label{Eq:d-gap2}
\end{eqnarray}

\begin{table}[tbp]
\centering
	\begin{tabular}{|c|c|c|}
	\hline
	angular momentum, $\ell$ & 2 & 4  \\
	\hline 
	$\frac{\Delta_{F}^{(\ell)}(T=0)}{\epsilon_{F}}$ & 0.406 & 0.197 \\
	\hline
	$\frac{k_B T_{c}^{(\ell)}}{\epsilon_{F}}$ & 0.185 & 0.090 \\
	\hline
	$\frac{2\Delta_{F}^{(\ell)}(T=0)}{k_B T_{c}}$ & 4.390 & 4.378 \\ 
	\hline
	\end{tabular}
\caption{The zero temperature gaps and transition temperatures for the angular momentum channels $\ell=2,4$.}
\label{Tab:gap}
\end{table}

\begin{figure}
	\centering
	\includegraphics[width=0.45\linewidth]{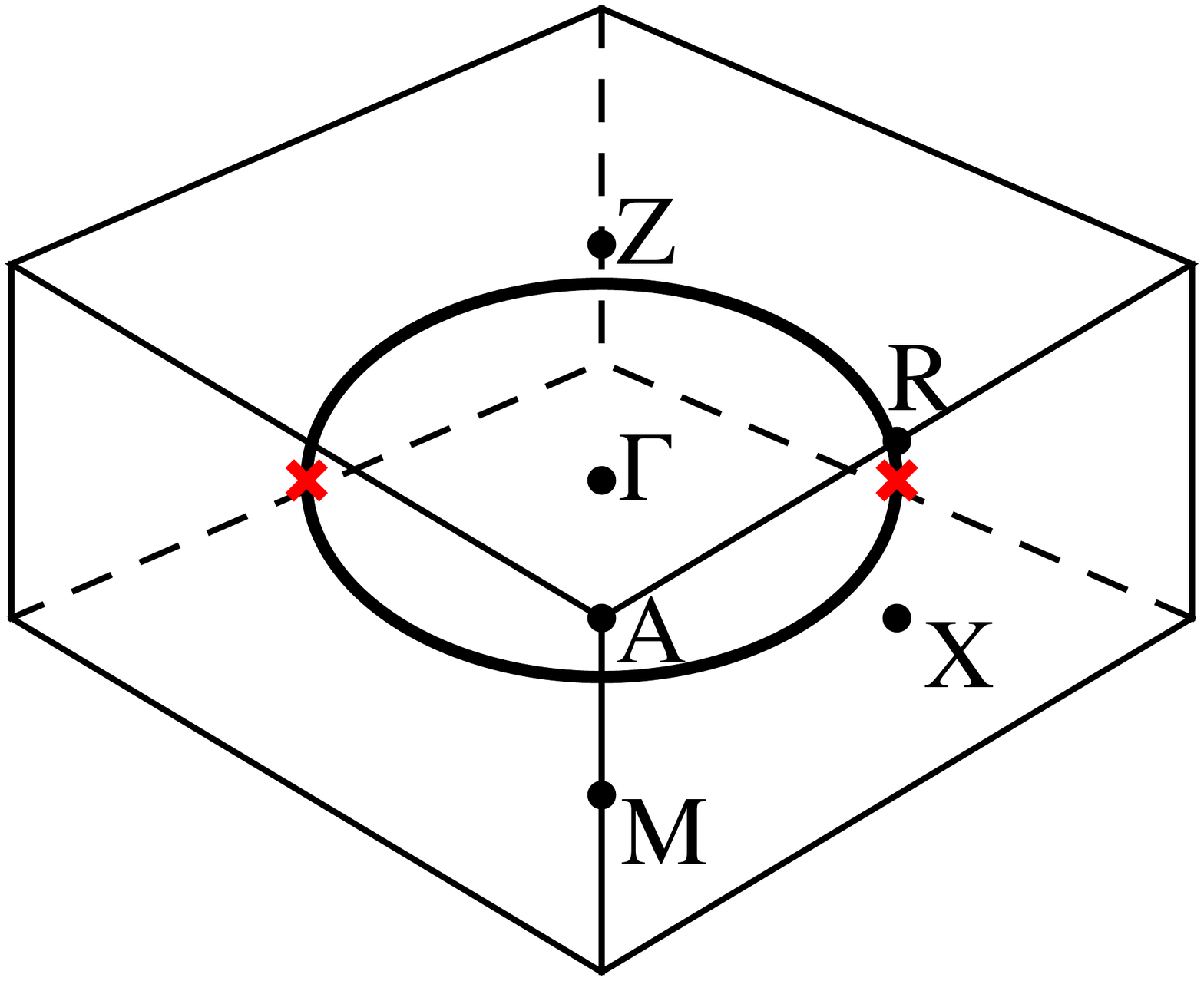}\hspace{0.05\linewidth} 
	\includegraphics[width=0.45\linewidth]{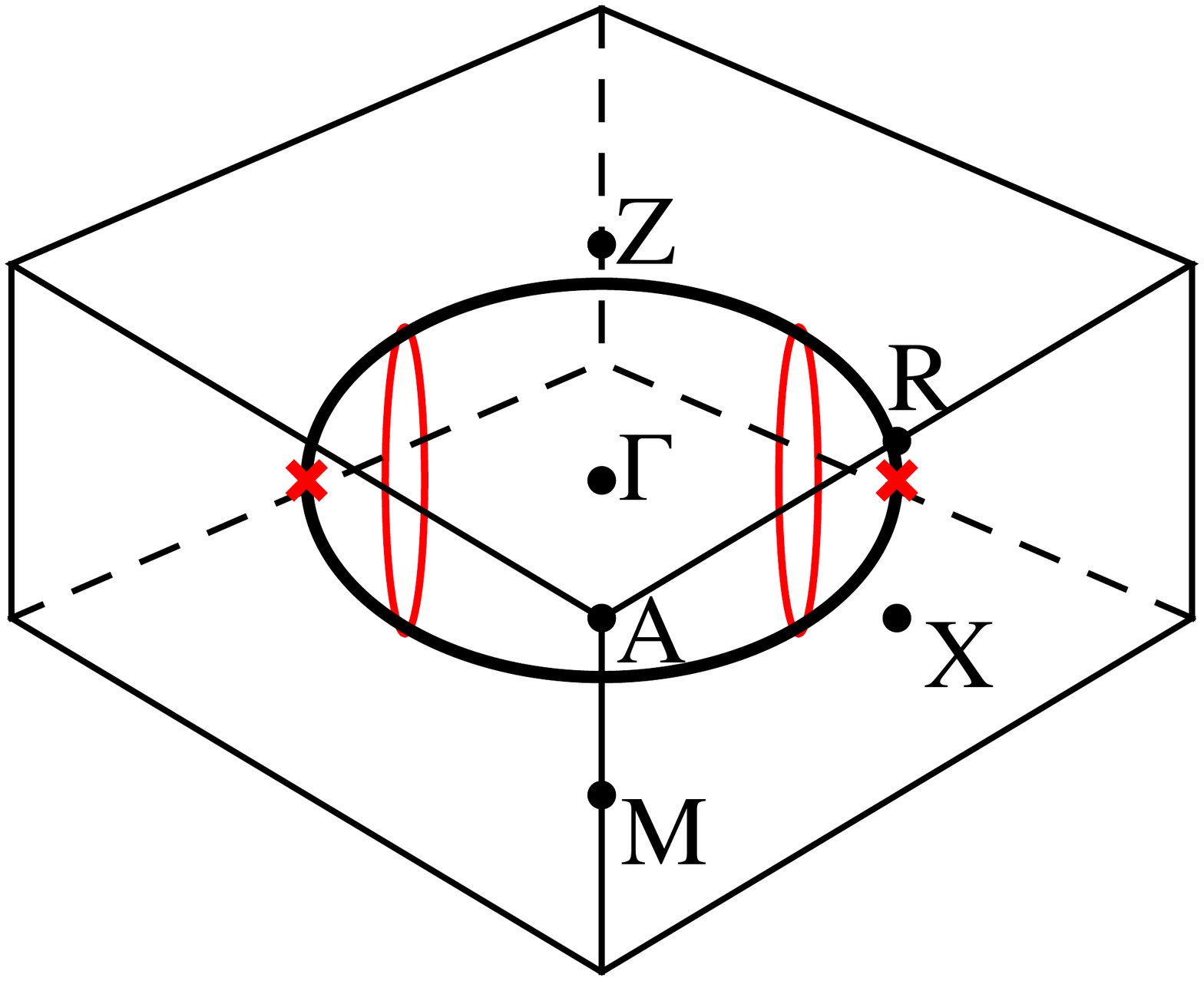}
	\caption{A cartoon illustration of possible nodal structures  drawn in the simple tetragonal unit cell due to the halving of the Brillouin zone because of  the nesting vector $(0,0,\frac{2\pi}{c})$.
The ellipsoids are  sketches of the Fermi surface. The red lines and crosses indicate the line and point nodes, respectively.  In the original coordinate the st-DDW order is on the diagonal planes in a manner consistent with the symmetries mentioned above.
Left: the point nodes of the pairing function in Eq.~(\ref{Eq:d-gap1}). Right: the line and point nodes of the pairing function in Eq.~(\ref{Eq:d-gap2}).}
	\label{Fig:nodes}
\end{figure}

Although both Eqs.(\ref{Eq:d-gap1}) and (\ref{Eq:d-gap2}) recover Eq.(\ref{Eq:d-gap}) in the long-wavelength limit, they have different nodal structures. The pairing function (\ref{Eq:d-gap1}) has only point nodes on the $k_{y'}$ axis (as $k_{x'}=k_{z'}=0$), but no line nodes. On the other hand, the pairing function (\ref{Eq:d-gap2}) has the same point nodes as (\ref{Eq:d-gap1}), as well as line nodes on the $k_{y'}a'=\pm \pi/2$ planes (FIG.~\ref{Fig:nodes}). The locations of the nodes are not the same as those proposed in Ref.~\citenum{Kasahara:2007}. However, the thermal conductivity measurements, which essentially probe the low-energy density of states, cannot unambiguously determine the location of the nodes in the momentum space. Thus, the order parameters (\ref{Eq:d-gap1}) and (\ref{Eq:d-gap2}) also adequately describe the low-temperature thermodynamic properties of the superconducting state based on the available experiments.
Notice that in order to compare the lattice to the continuum case, one needs to rescale the anisotropic lattice constants $a'$ and $c$ in the diagonal planes; i.e. to rescale the pink rectangles in FIG.~\ref{Fig:Pattern} into squares. 

\section{\label{Sec:discussion}Discussion}

Notice that the meron-antimeron pair which constitutes a skyrmion is not the same as the meron-meron Cooper pair as shown in FIG.~\ref{Fig:deconfine}. This is the reason why a skyrmion has zero angular momentum~\cite{Hsu:2013} while a Cooper pair formed by the merons may have nonzero angular momentum. Our theory is consistent with quantum oscillation measurements, carried out at milliKelvin temperature and high enough field to destroy superconductivity, once the three-dimensional character of the structure is taken into account. Of course, our simple band structure cannot correctly obtain all the observed frequencies.~\cite{Elgazzar:2009}

In three dimensions, one possible pairing function (\ref{Eq:d-gap1}) gives only point nodes in agreement with the specific heat measurements,~\cite{Yano:2008} whereas another possible pairing function (\ref{Eq:d-gap2}) has both line nodes and point nodes, as inferred from the thermal conductivity measurements.~\cite{Kasahara:2007} The actual nodal structure will ultimately be determined by experiments, which, to date, have not been able to rule out any of the possibilities of the pairing functions (\ref{Eq:d-gap1}) and (\ref{Eq:d-gap2}). Of course, there is also the possibility of replacing $\cos ( k_{y'} a' )$ in (\ref{Eq:d-gap2}) with other cosine functions such as $\cos \left( k_{z'} c \right)$ and $\cos \left( k_{y'} a'/2 \right)$, which also give Eq.(\ref{Eq:d-gap}) in the long-wavelength limit. However, without the knowledge of the actual Fermi surface and available momentum-sensitive experiments that would reveal the locations of the nodes, the pairing function cannot be unambiguously determined.

A rough  dimensional estimate of the magnitude of the PKE, if present, can be obtained from the imaginary part of the Hall conductivity, $\sigma_{xy}''\sim \frac{e^{2}}{h} (\frac{\Delta}{\Omega})^{2}$, where $\Omega$ is the laser frequency ($\sim 0.8 eV$). Broken TRS  is a necessary condition for  a nonzero  PKE.~\cite{Halperin:1992} Barring unknown effects, it is very likely that the experiments of PKE in the superconducting state is a reflection of the TRS breaking of our chiral order parameter.  Clearly there is room for much future improvements, but PKE~\cite{Kapitulnik:2009} can establish the nature of the superconducting state, a chiral $d$-wave condensate made out of fractionalized particles. This is consistent  with the latest measurements of Kapitulnik and his collaborators in the superconducting state.~\cite{Kapitulnik}
In their  measurements, the signals in the HO state become weaker in  samples with higher quality while those in the superconducting state appear to be robust. The signals in the superconducting state can be trained by magnetic field independently of the HO state, indicating that they have different origins. Moreover, the magnitude of the HO signals depend on the training field, and arise above $T_{\textrm{HO}}$ when the training field is large. Therefore, it is  likely that  PKE in the HO state is extrinsic, whereas TRS is actually broken in the superconducting state. 

In addition, in  PKE measurements, there is an anomaly at $T \sim$ 0.8-1.0~K within the superconducting phase. It is tempting to suggest that the subdominant order in $\ell=4$ channel is excited by the large  laser frequency. This could be very similar to $\mathrm{^{3}He}$ where the subdominant pairing in the $f$-wave channel is visible  only in  collective mode measurements.~\cite{Davis:2008,Sauls:1986} It would be interesting to vary the laser frequency, if possible.
 A less direct measurement of broken TRS in the superconducting state was recently presented in Ref.~\citenum{Li:2013}. On the other hand, a NMR experiment~\cite{Takagi:2012} finds conflicting results of  broken TRS in the HO state itself. There appear to be no data below $5 K$. So, we do not know how the signature of the broken TRS found in the HO state  relates to that below the superconducting transition temperature below $1.5 K$.
 Clearly further experiments will be necessary.
\begin{acknowledgments}
We thank E. Abrahams, S. B. Chung, W. P. Halperin, H.-Y. Kee, S. Kivelson, S. Raghu and S. Sachdev for their comments. We are also grateful to A. Kapitulnik and E. Schemm for the discussion on their PKE measurements. This work was supported by NSF under Grant No. DMR-1004520.
\end{acknowledgments}

\appendix
\section{Topological invariant}
\label{Topology}
The topology of the system can be studied by computing the Berry phase of the eigenstates. We define the Berry curvature, $\vec{\Omega}_{\sigma,\pm}$, as
\begin{equation}
  \vec{\Omega}_{\sigma,\pm} \equiv \vec{\bigtriangledown}_k \times \langle \Phi_{\sigma,\pm}(k)| i \vec{\bigtriangledown}_k | \Phi_{\sigma,\pm}(k) \rangle,
\end{equation}
where $|\Phi_{\sigma,\pm}(k) \rangle$ are the corresponding eigenstates. The Berry phase will be the integral of the Berry curvature over the reduced Brillouin zone (RBZ). 

Since the mixed st-DDW order is on the diagonal planes, one may expect that the nonzero contribution to Berry phase arises from the component of the Berry curvature perpendicular to the diagonal planes. This is indeed the case, and in order to simplify the calculation, we first rotate the coordinate along $z$ axis by $45^{\circ}$, i.e. $x'=(x-y)/\sqrt{2}$, $y'=(x+y)/\sqrt{2}$, and $z'=z$, 
\begin{eqnarray}
\epsilon_{1k} &=& 4t \left[ \cos \left( \frac{k_{x'} a'}{2} \right) + \cos \left( \frac{k_{y'} a'}{2} \right) \right] \cos \left( \frac{k_{z'} c}{2} \right), \\
\epsilon_{2k} &=& 4t' \cos \left( \frac{k_{x'} a'}{2} \right) \cos \left( \frac{k_{y'} a'}{2} \right) \nonumber \\
&& +2t^{\prime\prime} \left[
\cos \left( k_{x'} a' \right) + \cos \left( k_{y'} a' \right)
\right], \\
W_k        &=& W_0 \sin \left( \frac{k_{x'} a'}{2} \right) \sin \left( \frac{k_{z'} c}{2} \right), \\
\Delta_k   &=& \frac{\Delta_0}{2} \left[ \cos \left( k_{x'} a'\right) 
- \cos \left( k_{z'} c \right) \right]
\end{eqnarray}
with $a'=\sqrt{2}a$. The crystal structure is shown in FIG.~\ref{Fig:Pattern} in the main text, where we can find the primitive vectors in the rotated coordinate are $\frac{a'}{2}(\hat{x}'+\hat{y}')$, $\frac{a'}{2}(-\hat{x}'+\hat{y}')$ and $c\hat{z}'$. Therefore, the RBZ is bounded by $|k_{x'}a' \pm k_{y'}a'| = 2\pi$ and $|k_{z'}c| = \pi$. The spin-current patterns are now in the $x'z'$ planes ($y'=$ constant). We choose the spin quantization axis to be $y'$ axis, so $\sigma=\pm 1$ means the spin is along $\pm \hat{y}'$ direction. We will see below the nonzero contribution to the Berry phase is only from the $y'$ component. 

The mean-field st-DDW Hamiltonian can be written as
\begin{equation}
\mathcal{H}_{\textrm{st-DDW}}=\sum_{k,\sigma} (c_{k,\sigma}^{\dagger}, c_{k+Q,\sigma}^{\dagger}) \left( \epsilon_{2k} \tau^{0} + \vec{h}_{\sigma} \cdot \vec{\tau} \right) 
\left( 
\begin{array}{c}
c_{k,\sigma} \\
c_{k+Q,\sigma}
\end{array}
\right),
\end{equation}
where $\tau^{0}$ is 2 $\times$ 2 identity matrix and $\vec{\tau}$ are the Pauli matrices acting on the two-component spinor. Here the pseudospin vector is defined as $\vec{h}_{\sigma} \equiv (\Delta_k,-\sigma W_k,\epsilon_{1k} )$.
We have shown that the Berry curvature can be written in terms of the pseudospin vector $\vec{h}_{\sigma}$.~\cite{Hsu:2011} The $x'$, $y'$ and $z'$ components of the Berry curvature have the following forms,
\begin{eqnarray}
 (\Omega_{\sigma,\pm})_{x'} 
 &=& \mp  {\frac {1}{2E_k^3}} \vec{h}_{\sigma} \cdot ( {\frac {\partial{\vec{h}_{\sigma}}}{\partial k_{y'}}} \times {\frac {\partial{\vec{h}_{\sigma}}}{\partial k_{z'}}}), \\
 (\Omega_{\sigma,\pm})_{y'} 
 &=& \mp  {\frac {1}{2E_k^3}} \vec{h}_{\sigma} \cdot ( {\frac {\partial{\vec{h}_{\sigma}}}{\partial k_{z'}}} \times {\frac {\partial{\vec{h}_{\sigma}}}{\partial k_{x'}}}), \\
 (\Omega_{\sigma,\pm})_{z'} 
 &=& \mp  {\frac {1}{2E_k^3}} \vec{h}_{\sigma} \cdot ( {\frac {\partial{\vec{h}_{\sigma}}}{\partial k_{x'}}} \times {\frac {\partial{\vec{h}_{\sigma}}}{\partial k_{y'}}}). 
\end{eqnarray}

Notice that the Berry curvature does not depend on the $t',t''$ terms. To be explicit, we have
\begin{eqnarray}
 (\Omega_{\sigma,\pm})_{x'} 
&=& \mp {\frac {(-\sigma)}{2E_k^3}}
 \left| \begin{array}{ccc}
  \Delta_k & W_k & \epsilon_{1k} \vspace{0.1in} \\
  {\frac {\partial{\Delta_k}}{\partial k_{y'}}} & {\frac {\partial{W_k}}{\partial k_{y'}}}&{\frac {\partial{\epsilon_{1k}}}{\partial k_{y'}}} \vspace{0.1in} \\
  {\frac {\partial{\Delta_k}}{\partial k_{z'}}} & {\frac {\partial{W_k}}{\partial k_{z'}}}&{\frac {\partial{\epsilon_{1k}}}{\partial k_{z'}}} \vspace{0.1in}
  \end{array}\right| \nonumber \\
&=& \pm \sigma \frac{tW_0\Delta_0a'c}{ 4E_k^3} \sin \left( \frac{k_{x'}a'}{2}\right) \sin \left( \frac{k_{y'}a'}{2}\right)  \nonumber \\
&&
\times \cos^2 \left( \frac{k_{z'}c}{2}\right) \left[ -2 + \cos \left( k_{x'}a'\right) + \cos \left( k_{z'}c\right) \right] \nonumber \\
\end{eqnarray}

For the $y'$ component, we have
\begin{eqnarray}
 (\Omega_{\sigma,\pm})_{y'} 
&=& \mp {\frac {(-\sigma)}{2E_k^3}}
 \left| \begin{array}{ccc}
  \Delta_k & W_k & \epsilon_{1k} \vspace{0.1in} \\
  {\frac {\partial{\Delta_k}}{\partial k_{z'}}} & {\frac {\partial{W_k}}{\partial k_{z'}}}&{\frac {\partial{\epsilon_{1k}}}{\partial k_{z'}}} \vspace{0.1in} \\
  {\frac {\partial{\Delta_k}}{\partial k_{x'}}} & {\frac {\partial{W_k}}{\partial k_{x'}}}&{\frac {\partial{\epsilon_{1k}}}{\partial k_{x'}}} \vspace{0.1in}
  \end{array}\right| \nonumber \\
&=& \pm \sigma \frac{tW_0\Delta_0a'c}{ 8E_k^3} \left[ 2 - \cos \left( k_{x'}a'\right) - \cos \left( k_{z'}c\right) \right] \nonumber \\
&& 
\times \left\{ 2 + \cos \left( k_{x'}a'\right)  + \cos \left( k_{z'}c\right) \right. \nonumber \\
&&
\left. \hspace{0.1in} + \cos \left( \frac{k_{x'}a'}{2}\right) \cos \left( \frac{k_{y'}a'}{2} \right) 
\left[ 3 +  \cos \left( k_{z'}c\right)  \right] \right\}  \nonumber \\
\end{eqnarray}

For the $z'$ component, we have
\begin{eqnarray}
 (\Omega_{\sigma,\pm})_{z'} 
&=& \mp {\frac {(-\sigma)}{2E_k^3}}
 \left| \begin{array}{ccc}
  \Delta_k & W_k & \epsilon_{1k} \vspace{0.1in} \\
  {\frac {\partial{\Delta_k}}{\partial k_{x'}}} & {\frac {\partial{W_k}}{\partial k_{x'}}}&{\frac {\partial{\epsilon_{1k}}}{\partial k_{x'}}} \vspace{0.1in} \\
  {\frac {\partial{\Delta_k}}{\partial k_{y'}}} & {\frac {\partial{W_k}}{\partial k_{y'}}}&{\frac {\partial{\epsilon_{1k}}}{\partial k_{y'}}} \vspace{0.1in}
  \end{array}\right| \nonumber \\
&=& \pm \sigma \frac{tW_{0}\Delta_{0} a^{\prime 2}}{ 8E_k^3} \cos \left( \frac{k_{x'}a'}{2}\right) \sin \left( \frac{k_{y'}a'}{2}\right) \nonumber \\ 
&& \times \sin \left( k_{z'}c \right) \left[ -2 + \cos \left( k_{x'}a'\right) + \cos \left( k_{z'}c\right) \right] \nonumber \\
\end{eqnarray}

Although $(\Omega_{\sigma,\pm})_{x'}$ and $(\Omega_{\sigma,\pm})_{z'}$ are nonzero, their integrals over the RBZ are zero.
In other words, because of the planar structure of the order parameter, the Berry curvature and the topology will be similar to the two-dimensional model~.\cite{Hsu:2011}
Under time reversal, the $k_{y'}a'=\pi$ plane maps onto the $k_{y'}a'=-\pi$ plane, which is equivalent to the $k_{y'}a'=\pi$ plane because there is a reflection symmetry under $k_{y'} \leftrightarrow - k_{y'}$.
Therefore, the $k_{y'}a'=\pi$ plane maps onto itself under time reversal, so there is a topological invariant associated with the $k_{y'}a'=\pi$ plane.
On the other hand, the $k_{y'}a'=0$ plane, which also maps onto itself under time reversal, is not fully gapped, so topological invariant is not well defined on this plane. In addition, for $k_{y'}a'=\pi$, the spectrum is fully gapped, which leads to an exponential behavior of the specific heat consistent with experiments and therefore implies that the relevant physics is on the $k_{y'}a'=\pi$ plane. 
 
For the $y'$ component, we thus project the Berry curvature to the $k_{y'}a'=\pi$ plane and perform the integral. The Chern number for each band will be
\begin{eqnarray}
 N_{\sigma,\pm} &=& \frac{1}{2\pi} \int_{-\frac{\pi}{a'}}^{\frac{\pi}{a'}} dk_{x'} \int_{-\frac{\pi}{c}}^{\frac{\pi}{c}} dk_{z'} (\Omega_{\sigma,\pm})_{y'}  \nonumber \\
&=& \pm \sigma,
\end{eqnarray}
which does not depend on the details of the band parameters as long as $t,W_{0}$, and $\Delta_{0}$ are all nonzero. Therefore, the topology of the system is nontrivial, and there will be a quantized spin Hall conductance on the $x'z'$ planes, $\sigma_{x'z'}^{\textrm{spin}} = {\frac {e}{2\pi}} $.

\section{Skyrmions in the system}
\subsection{Low-energy action}

As mentioned above, because of the planar structure of the order parameter, the interesting physics, such as the spin current patterns and the quantized spin Hall conductance, will be on the diagonal $x'z'$ planes ($y'=$ constant). So we may project the system onto the $x'z'$ planes to construct the linearized low-energy action. Again, the spin quantization axis is along $y'$ axis, which is perpendicular to the $x'z'$ planes. 

Defining the spinor $\psi_{k,\alpha}^{\dagger} \equiv (c_{k,\alpha}^{\dagger}, c_{k+Q,\alpha}^{\dagger})$ with spin index $\alpha$ (not to be confused with the meron operator in the main text), the Hamiltonian can be written as
\begin{eqnarray}
\mathcal{H}_{\textrm{st-DDW}} &=& 
\sum_{\alpha,\beta} \sum_{k} \psi_{k,\alpha}^{\dagger} 
\left[  \delta_{\alpha\beta} \tau^{z} \epsilon_{1k} +
 \delta_{\alpha\beta} \tau^{x} \Delta_k \right. \nonumber \\
&& \hspace{0.7in} \left.
- (\vec{\sigma} \cdot \hat{N})_{\alpha\beta} \tau^{y} W_k \right]  \psi_{k,\beta},
\end{eqnarray}
where $\tau^{i}$ $(i=x,y,z$) are Pauli matrices acting on the two-component spinor. For simplicity we have set $t'=t''=\mu=0$ and $k_{y'}a'=\pi$. Then, we can construct the low-energy effective model by linearizing the action around the following points,
\begin{eqnarray}
\vec{K}_1 &\equiv& \frac{\pi}{c} \hat{z}', \\
\vec{K}_2 &\equiv& \frac{\pi}{a'} \hat{x}', \\
\vec{K}_3 &\equiv& \frac{\pi}{a'} \hat{x}' + \frac{\pi}{c} \hat{z}',
\end{eqnarray} 
where $\vec{K}_1$ and $\vec{K}_2$ are the nodal points in the absence of the singlet component ($\Delta_0=0$), and $\vec{K}_3$ is the nodal point in the absence of the triplet component ($W_0=0$).

Therefore, the linearized low-energy action will be
\begin{widetext}
\begin{eqnarray}
S &=& \int d^3 x  \left\{
\psi_{1}^{\dagger} 
\left[ -\partial_{\tau} +2t \tau^{z} c (\frac{1}{i}\partial_{z'}) 
- \tau^{x} \Delta_0 
+ (\vec{\sigma} \cdot \hat{N}) \tau^{y} \frac{W_0}{2} a' (\frac{1}{i}\partial_{x'}) \right] 
 \psi_{1} \right. \nonumber \\
&& \hspace{0.5in} + \psi_{2}^{\dagger}
\left[ -\partial_{\tau}  +2t \tau^{z} a' (\frac{1}{i}\partial_{x'}) 
+ \tau^{x} \Delta_0 
+ (\vec{\sigma} \cdot \hat{N}) \tau^{y} \frac{W_0}{2} c (\frac{1}{i}\partial_{z'}) 
\right] 
 \psi_{2}  
 \nonumber \\
&& \hspace{0.5in} \left.  
+ \psi_{3}^{\dagger} 
\left[ -\partial_{\tau}
+(\vec{\sigma} \cdot \hat{N}) \tau^{y} W_0 \right] \psi_{3}
\right\},
\end{eqnarray}
\end{widetext}
where we have introduced the imaginary time $i\partial_{t} = - \partial_{\tau}$. 

Notice that there is no spatial derivative in the $\psi_{3}$ term since the expansion of the form factor $W_{k}$ around the nodal point $\vec{K}_{3}$ is 
\begin{equation}
W_{K_3+q} = W_0 (1-\frac{q_{x'}^2a^{\prime 2}}{8}-\frac{q_{z'}^2c^2}{8} +\cdots),
\end{equation}
where the second- (and higher-) order derivative terms are dropped when linearizing the action. In other words, the $W_{k}$ term behaves as a mass term at the $\vec{K}_3$ point.

\subsection{The charges of the skyrmions: an adiabatic argument}

We will compute the charge of skyrmions in the system by the adiabatic argument~\cite{Grover:2008}. Consider the action around $\vec{K}_1$ when the order parameter is uniform (say, $\hat{N}=\hat{y}'$). The results for $\vec{K}_2$ and $\vec{K}_3$ follow identically.

\begin{eqnarray}
S_{1} &=& \int d^3 x \psi_{1}^{\dagger} 
\left[ -\partial_{\tau} +2t \tau^{z} c (\frac{1}{i}\partial_{z'}) 
- \tau^{x} \Delta_0 \right. \nonumber \\
&& \hspace{0.6in} \left.
+ \sigma^{y'} \tau^{y} \frac{W_0}{2} a' (\frac{1}{i}\partial_{x'}) 
 \right] \psi_{1} 
\end{eqnarray}

As mentioned above, the spin quantization axis is now along $y'$ axis, so $\sigma=\pm 1$ means the spin is along $\pm \hat{y}'$ direction, and the  Pauli spin matrices are 
\begin{eqnarray}
\sigma^{x'} = \left(
\begin{array}{cc}
0 & -i \\
i & 0
\end{array}
\right); \;
&
\sigma^{y'} = \left(
\begin{array}{cc}
1 & 0 \\
0 &-1
\end{array}
\right); \;
&
\sigma^{z'} = \left(
\begin{array}{cc}
0 & 1 \\
1 & 0
\end{array}
\right) \nonumber \\
\end{eqnarray}

In the previous section we have shown that the nontrivial topology leads to a quantized spin Hall conductance on the $x'z'$ planes. The quantized spin Hall conductance implies that the external gauge fields $A^c$ and $A^s$ couple to spin and charge currents, respectively. In the presence of these external gauge fields, we add minimal coupling in the action by taking
\begin{eqnarray}
\frac{1}{i} \partial_{\mu} \rightarrow  \frac{1}{i} \partial_{\mu} +A^c_{\mu} + \frac{\sigma^{y'}}{2} A^s_{\mu},
\end{eqnarray}
and the action can be written as
\begin{widetext}
\begin{eqnarray}
S_{1}[A^c,A^s] &=& \int d^3 x \psi_{1}^{\dagger} 
\left[ -i \left( \frac{1}{i}\partial_{\tau} +A^c_{\tau} + \frac{\sigma^{y'}}{2} A^s_{\tau} \right) +2t \tau^{z} c \left( \frac{1}{i} \partial_{z'} +A^c_{z'} + \frac{\sigma^{y'}}{2} A^s_{z'} \right) \right. \nonumber \\
&& \hspace{0.6in} \left.
- \tau^{x} \Delta_0 
+ \sigma^{y'} \tau^{y} \frac{W_0}{2} a' \left( \frac{1}{i} \partial_{x'} +A^c_{x'} + \frac{\sigma^{y'}}{2} A^s_{x'} \right) 
 \right] \psi_{1}, 
\label{Eq:min}
\end{eqnarray}
\end{widetext}
where we set $e=\hbar=c=1$. The non-vanishing transverse spin conductance implies that the low-energy effective action for the gauge fields is given by
\begin{eqnarray}
S_{1,\textrm{eff}}=\frac{i}{2\pi} \int d^3 x \epsilon^{\mu\nu\lambda}A^c_{\mu}\partial_{\nu}A^s_{\lambda},
\end{eqnarray}
and the charge current is induced by the spin gauge field
\begin{eqnarray}
j^c_{\mu}= \frac{1}{2\pi} \epsilon^{\mu\nu\lambda}\partial_{\nu}A^s_{\lambda}.
\end{eqnarray}
Notice that the prefactor comes from the quantized spin Hall conductance  $\sigma_{x'z'}^{\textrm{spin}} = {\frac {e}{2\pi}} $, so this is a physical charge current. 

Consider a static configuration of the $\hat{N}$ field with Pontryagin index one,
\begin{equation}
\hat{N}(r,\theta)=\left[ \sin \alpha(r) \sin \theta, \cos \alpha(r), \sin \alpha(r) \cos \theta \right],
\label{Eq:config}
\end{equation}
where $(r,\theta)$ is the polar coordinate defined on the $x'z'$ planes and $\alpha(r)$ satisfies the boundary conditions $\alpha(r=0)=0$ and $\alpha(r\rightarrow \infty)=\pi$. This field configuration corresponds to one skyrmion, and now the action is
\begin{eqnarray}
S_{1} &=& \int d^3 x \psi_{1}^{\dagger} 
\left[ -\partial_{\tau} +2t \tau^{z} c (\frac{1}{i}\partial_{z'}) 
- \tau^{x} \Delta_0 \right. \nonumber \\
&& \hspace{0.6in} \left.
+ (\vec{\sigma} \cdot \hat{N}) \tau^{y} \frac{W_0}{2} a' (\frac{1}{i}\partial_{x'}) 
\right] \psi_{1} 
\label{Eq:Seff_1}
\end{eqnarray}

We can perform a unitary transformation at all points in space such that
\begin{equation}
U^{\dagger} (\vec{\sigma} \cdot \hat{N}) U = \sigma^{y'}
\label{Eq:unitary}
\end{equation}

Defining $\psi=U \psi'$, and plugging into Eq.(\ref{Eq:Seff_1}), we obtain
\begin{widetext}
\begin{eqnarray}
S_{1} &=& \int d^3 x \psi_{1}^{\prime\dagger} U^{\dagger}
\left[ -\partial_{\tau} +2t \tau^{z} c (\frac{1}{i}\partial_{z'}) 
- \tau^{x} \Delta_0 
+ (\vec{\sigma} \cdot \hat{N}) \tau^{y} \frac{W_0}{2} a' (\frac{1}{i}\partial_{x'}) 
 \right] U\psi'_{1} \nonumber \\
 &=& \int d^3 x \psi_{1}^{\prime\dagger} 
\left[  -\partial_{\tau} +2t \tau^{z} c (\frac{1}{i}\partial_{z'}) 
- \tau^{x} \Delta_0 
+ \sigma^{y'} \tau^{y} \frac{W_0}{2} a' (\frac{1}{i}\partial_{x'}) 
 \right] \psi'_{1}  \nonumber\\
 && +\int d^3 x \psi_{1}^{\prime\dagger} 
\left[  -\left( U^{\dagger}\partial_{\tau}U\right) +2t \tau^{z} c \left( \frac{1}{i}U^{\dagger}\partial_{z'}U\right) 
+ \sigma^{y'} \tau^{y} \frac{W_0}{2} a' \left(\frac{1}{i}U^{\dagger}\partial_{x'}U\right) 
 \right] \psi'_{1} 
\label{Eq:rot}
\end{eqnarray}
\end{widetext}

Equating Eq.(\ref{Eq:rot}) and Eq.(\ref{Eq:min}), we have
$A^c_{\tau} = A^s_{\tau} = A^c_{x'} = A^c_{z'} =0$, and
\begin{eqnarray}
\frac{1}{i}  U^{\dagger} \partial_{x'} U &=&   \frac{\sigma^{y'}}{2} A^s_{x'}, \\
\frac{1}{i}  U^{\dagger} \partial_{z'} U &=&   \frac{\sigma^{y'}}{2} A^s_{z'}. 
\end{eqnarray}

In the far-field limit ($r \rightarrow \infty$), the unitary matrix is
\begin{eqnarray}
U(r\rightarrow \infty,\theta)=\left(
\begin{array}{cc}
0 & - e^{-i\theta}\\
e^{i\theta} & 0
\end{array}
\right),
\end{eqnarray}
so we have 
\begin{eqnarray}
\vec{A}^s 
&=& -\frac{2 \sin \theta}{r} \hat{z}' + \frac{2 \cos \theta}{r} \hat{x}' 
= \frac{2}{r} \hat{\theta},
\end{eqnarray}
which is in the $x'z'$ planes.

In other words, threading a skyrmion into the system is equivalent to adding an external gauge field $\vec{A}^s$ with a flux of $4\pi$ in the $y'$ direction. Suppose we adiabatically construct the skyrmionic configuration $\hat{N}(r,\theta)$ from the ground state $\hat{y}'$ in a very large time period $\tau_p \rightarrow \infty$. During the process, we effectively thread a gauge flux of $4\pi$ adiabatically into the $x'z'$ planes. The quantized spin Hall conductance implies that a radial current  will be induced by the $4\pi$ gauge flux of $\vec{A}^s(t)$, which is now time-dependent: $\vec{A}^s(t=0)=0$ and $\vec{A}^s(t=\tau_p)=\vec{A}^s$. That is,
\begin{eqnarray}
j^c_{r}(t) = -\frac{1}{2\pi} \partial_t A^s_{\theta}(t).
\end{eqnarray}

As a result, charge will be transferred from the center to the boundary, and the total charge transferred during the process can be computed by performing the integral
\begin{eqnarray}
Q^c&=&\int^{\tau_p}_{0} dt \int^{2\pi}_{0} r d\theta j^c_r(t) \nonumber \\
&=& -2.
\end{eqnarray}
Therefore, we obtain a skyrmion with charge $2e$ and flux $4\pi$, as in the two-dimensional model~\cite{Hsu:2013}.

One can also infer the existence of skyrmions by integrating out the fermions and thereby formulating a non-linear $\sigma$-model.~\cite{Grover:2008} Here its is 
a lengthy and tedious exercise and is therefore not reproduced.
\begin{section}{The superconducting gap}
\label{Appendix:D}
\begin{widetext}
To save space, here we only write down the equations explicitly for $T=0$. For finite temperatures one merely has to substitute 
\begin{equation}
\frac{\Delta_{k'}^{sc}(T)}{2\sqrt{\xi_{k'}^2 + |\Delta_{k'}^{sc}(T)|^2 }} \to \frac{\Delta_{k'}^{sc}(T)}{2\sqrt{\xi_{k'}^2 + |\Delta_{k'}^{sc}(T)|^2 }} \times \tanh \left(
\frac{1}{2k_{B} T} \sqrt{\xi_{k'}^2 + |\Delta_{k'}^{sc}(T)|^2 } \right)
\end{equation}
We begin with the ansatz for $l$ wave pairing (for brevity, we shall drop the superscript $sc$ on the gap $\Delta$,without confusion, we hope),
\begin{equation}
\Delta_k = |\Delta_k| e^{i l \phi_k},
\end{equation}
where $\phi_k$ denotes the direction of the wave vector, and we will choose it to be the angle between $\vec{k}$ and $\vec{k}'$ for simplicity.
Plugging the ansatz into the gap equation, we have~\cite{Greiter:1992,Morinari:2006}
\begin{eqnarray}
|\Delta_k| 
&=& -  \frac{2\pi i}{m} \int \frac{d^2k'}{(2\pi)^2}
\frac{(\vec{k} \times \vec{k}')_z}{|\vec{k} - \vec{k}'|^2}
\frac{\Delta_{k'}}{\sqrt{\xi_{k'}^2 + |\Delta_{k'}|^2 }} \nonumber\\
&=&  -  \frac{ i}{2\pi m} \int_0^{\infty} k' dk' \int_0^{2\pi} d\phi
\frac{k k' \sin \phi }{k^2+k'^2-2kk'\cos \phi}
\frac{|\Delta_{k'}|e^{il\phi}}{\sqrt{\xi_{k'}^2 + |\Delta_{k'}|^2 }} \nonumber\\
&=&  -  \frac{ i}{2\pi m} \int_0^{\infty} k' dk' 
\frac{|\Delta_{k'}|}{\sqrt{\xi_{k'}^2 + |\Delta_{k'}|^2 }} 
\int_0^{2\pi} d\phi
\frac{k k' \sin \phi e^{il\phi}}{k^2+k'^2-2kk'\cos \phi}  \nonumber\\
&=&  -  \frac{ i}{4\pi m} \int_0^{\infty} k' dk' 
\frac{|\Delta_{k'}|}{\sqrt{\xi_{k'}^2 + |\Delta_{k'}|^2 }} 
\int_0^{2\pi} d\phi
\frac{\sin \phi e^{il\phi}}{ \lambda_{kk'} -\cos \phi},
\end{eqnarray}
where $\lambda_{kk'} \equiv \frac{k^2+k'^2}{2kk'}$.
\end{widetext}
The angular integral can be computed  by performing a contour integral in the complex plane. To do this, we set $z=e^{i\phi}$ and get
\begin{eqnarray}
I_l(\lambda) 
&\equiv& \int_0^{2\pi} d\phi
\frac{\sin \phi e^{il\phi}}{ \lambda -\cos \phi}  \nonumber\\
&=& \int_0^{2\pi} d\phi
\frac{ (\frac{e^{i\phi}-e^{-i\phi}}{2i}) e^{il\phi}}{ \lambda -(\frac{e^{i\phi}+e^{-i\phi}}{2})} \nonumber\\
&=& \oint \frac{dz}{iz} \frac{1}{i} 
\frac{ (z-z^{-1}) z^{l}}{ 2\lambda -(z+z^{-1})} \nonumber\\
&=& -\oint \frac{dz}{z} 
\frac{ (z^2-1) z^{l}}{ 2\lambda z - z^2 -1 },
\end{eqnarray}
where we have used $d \phi = \frac{dz}{iz}$ and converted the $\phi$ integral into a contour integral around the origin with unit radius. 

\begin{eqnarray}
I_l(\lambda) 
&=& \oint dz 
\frac{ (z^2-1) z^{l-1}}{ z^2 - 2\lambda z + 1 } \nonumber\\
&=& \oint dz 
\frac{ (z^2-1) z^{l-1}}{ (z-\lambda-\sqrt{\lambda^2-1})(z-\lambda+\sqrt{\lambda^2-1}) },
\end{eqnarray}
where the poles are at $z=\lambda \pm \sqrt{\lambda^2-1}$. Since $z=\lambda + \sqrt{\lambda^2-1}\ge 1$ for $\lambda \ge 1$, it is not enclosed by the contour. Thus, only $z=\lambda - \sqrt{\lambda^2-1}$ contributes to the integral, and we get
\begin{eqnarray}
I_l(\lambda) 
&=& 2\pi i \left[
\frac{ (z^2-1) z^{l-1}}{ (z-\lambda-\sqrt{\lambda^2-1}) }
\right]_{z =\lambda - \sqrt{\lambda^2-1}} \nonumber \\
&=& 2\pi i 
(\lambda - \sqrt{\lambda^2-1})^{l-1}
\left[
\frac{ \lambda^2 + \lambda^2-1 - 2 \lambda \sqrt{\lambda^2-1} -1 }{ -2\sqrt{\lambda^2-1} }
\right] \nonumber \\
&=& 2\pi i 
(\lambda - \sqrt{\lambda^2-1})^{l}
\end{eqnarray}
\begin{widetext}
Since $\lambda_{kk'} \equiv \frac{k^2+k'^2}{2kk'}$, we have
\begin{eqnarray}
\lambda_{kk'} - \sqrt{\lambda_{kk'}^2-1} 
&=& \frac{k^2+k'^2}{2kk'} -
\sqrt{
\left( \frac{k^2+k'^2}{2kk'}+1\right) \left( \frac{k^2+k'^2}{2kk'}-1\right)
} \nonumber \\
&=& \frac{1}{2kk'} \left[
k^2+k'^2 - |k^2 - k'^2|
\right] \nonumber \\
&=& \left\{ 
\begin{array}{ll}
\left(\frac{k'}{k}\right) & \textrm{ as } k>k' \\
\left(\frac{k}{k'}\right) & \textrm{ as } k<k'
\end{array}
\right.
\end{eqnarray}
\end{widetext}
So, 
\begin{eqnarray}
I_l (\lambda_{kk'}) &=& 2\pi i \times \left\{ 
\begin{array}{ll}
\left(\frac{k'}{k}\right)^{l} & \textrm{ as } k>k' \\
\left(\frac{k}{k'}\right)^{l} & \textrm{ as } k<k'
\end{array}
\right.
\end{eqnarray}
\begin{widetext}
Putting back into the gap equation, we have
\begin{equation}
|\Delta_k| 
= \frac{ 1 }{2m} \left[
\int_0^{k}  dk' \left(\frac{k'}{k}\right)^{l}
\frac{k'|\Delta_{k'}|}{\sqrt{\xi_{k'}^2 + |\Delta_{k'}|^2 }}  
+\int_k^{\infty}  dk' \left(\frac{k}{k'}\right)^{l}
\frac{k'|\Delta_{k'}|}{\sqrt{\xi_{k'}^2 + |\Delta_{k'}|^2 }}  
\right]
\label{Eq:gap1}
\end{equation}
Hence we can see that $|\Delta_k| \propto k^{-l}$ as $k \rightarrow \infty$ and $|\Delta_k| \propto k^{l}$ as $k \rightarrow 0$. So we make the fallowing  ansatz, which solves the gap equation 
self-consistently,
\begin{eqnarray}
|\Delta_k^{(\ell)}| &=& \left\{ 
\begin{array}{ll}
\Delta_{F}^{(\ell)} \left(\frac{k_{F}}{k}\right)^{l} & \textrm{ as } k \ge k_{F}, \\ 
\Delta_{F}^{(\ell)}\left(\frac{k}{k_{F}}\right)^{l} & \textrm{ as } k \le k_{F},
\end{array}
\right.
\label{Eq:gap-ansatz}
\end{eqnarray}
with the Fermi wave vector $k_{F}$.  
Dividing both sides by $\epsilon_F$ and letting $u \equiv \frac{\Delta_F^{(\ell)}}{\epsilon_F} = \frac{2m\Delta_F^{(\ell)}}{k_F^2}$, we get
\begin{eqnarray}
u &=& 
\int_0^{1}  dx 
\frac{ u x^{2l+1} }{\sqrt{(x^2-1)^2 + u^2 x^{2l} }} 
+\int_{1}^{\infty}  dx 
\frac{ u x^{1-2l} }{\sqrt{(x^2-1)^2 + u^2 x^{-2l} }}
\end{eqnarray}
We can further simplify the equation by setting $y=x^2$ and obtain
\begin{eqnarray}
u &=& 
\frac{1}{2} \int_0^{1}  dy 
\frac{ u y^{l} }{\sqrt{(y-1)^2 + u^2 y^{l} }} 
+\frac{1}{2} \int_{1}^{\infty}  dy 
\frac{ u y^{-l} }{\sqrt{(y^2-1)^2 + u^2 y^{-l} }}
\label{Eq:gap2}
\end{eqnarray}
which is be solved by iterating
Eq.~(\ref{Eq:gap2})  using Mathematica,
\begin{eqnarray}
u_{out} &=& 
\frac{1}{2} \int_0^{1}  dy 
\frac{ u_{in} y^{l} }{\sqrt{(y-1)^2 + u_{in}^2 y^{l} }} 
+\frac{1}{2} \int_{1}^{\infty}  dy 
\frac{ u_{in} y^{-l} }{\sqrt{(y^2-1)^2 + u_{in}^2 y^{-l} }}.
\label{Eq:gap3}
\end{eqnarray}
\end{widetext}

\end{section}

\end{document}